\documentclass[11pt,a4paper]{article}
\usepackage{epsfig}
\usepackage{amsmath}
\usepackage{amssymb}
\usepackage{verbatim}
\usepackage{multirow}
\usepackage{bm}
\usepackage{dsfont}
\usepackage[T1]{fontenc}
\usepackage[footnotesize]{caption}
\usepackage{subfigure}
\usepackage{epsfig}
\usepackage{cite}
\usepackage{textcomp}
\usepackage{calc}
\usepackage[margin=0.9in]{geometry}
\usepackage{bbm}
\usepackage[dvipsnames]{xcolor}
\usepackage{color,graphicx}
\usepackage[utf8]{inputenc}
\usepackage{lipsum,tabularx}
\usepackage[usestackEOL]{stackengine}
\usepackage[margin=1cm]{caption}
\edef\tmp{\the\baselineskip}
\setstackgap{L}{\tmp}
\definecolor{pcolor}{RGB}{117,19,7}
\usepackage[colorlinks=true
  ,urlcolor=pcolor
  ,anchorcolor=pcolor
  ,citecolor=pcolor
  ,filecolor=blue
  ,linkcolor=pcolor
  ,menucolor=blue
  ,pagecolor=blue
  ,linktocpage=true
  ,pdfproducer=medialab
  ,pdfa=true
]{hyperref}

\begin{document}


\vspace*{1.3cm}
\vskip 1.5cm

\begin{center}

\bigskip
{\huge\bf  Oscillon collapse to black holes}

\vskip 1.2cm

{\large \bf
Zainab~Nazari$^{a,b}$,
 Michele~Cicoli$^{c,d}$,
 Katy~Clough$^e$, and
Francesco~Muia$^f$}\\[8mm]

{\it{
$^{a}${Abdus Salam International Centre for Theoretical Physics, 34151, Trieste, Italy}, \\
$^{b}${Fizik B\"{o}l\"{u}m\"{u}, Bo\u{g}azi\c{c}i  \"{U}niversitesi,  Bebek, 34342, Istanbul, Turkey} \\
$^c${Dipartimento di Fisica e Astronomia, Universit\`a di Bologna, via Irnerio 46, 40126 Bologna, Italy}, \\
$^d${INFN, Sezione di Bologna, viale Berti Pichat 6/2, 40127 Bologna, Italy} \\
$^e${Astrophysics, University of Oxford, DWB, Keble Road, Oxford OX1 3RH, UK} \\
$^f${DAMTP, Centre for Mathematical Sciences, Wilbeforce Road, Cambridge, CB3 0WA, UK}}}

\vskip 0.6cm

\it{\href{mailto:znazari@ictp.it}{znazari@ictp.it}, \href{mailto:michele.cicoli@unibo.it}{michele.cicoli@unibo.it},\\ \href{mailto:katy.clough@physics.ox.ac.uk}{katy.clough@physics.ox.ac.uk}, \href{mailto:fm538@damtp.cam.ac.uk}{fm538@cam.ac.uk}}

\vskip 1.0cm

\end{center}


\abstract{Using numerical relativity simulations we study the dynamics of pseudo-topological objects called oscillons for a class of models inspired by axion-monodromy. Starting from free field solutions supported by gravitational attractions, we investigate the effect of adding self-interactions, and contrast this with the effect of adding self-interactions whilst removing gravitational support. We map out regions of the parameter space where the initial conditions rapidly collapse to black holes, and other regions where they remain pseudo-stable or disperse.}

%

\tableofcontents

\section{Introduction}
\label{sec:Introduction}

Oscillons and similar objects have recently gained a considerable amount of attention due to their possible role in the early universe~\cite{1006.3075, 0712.3034, 1004.4658, astro-ph/9311037, 1002.3380, 1009.2505, 1103.1911, 1106.3335,  1710.06851, Antusch:2017vga, 1711.10496,  1902.07261, 1902.06736, Sang:2019ndv}. They are predicted to be formed in many post-inflationary scenarios, affecting the transition from inflation to the standard radiation dominated phase of the early universe. Most importantly, their formation typically leads to the production of gravitational waves with a peculiar spectrum, characterised by a peak centered around the value of the mass of the scalar field involved in the process~\cite{1304.6094, 1607.01314, 1707.09841, 1708.08922, 1803.08047}. Despite this peak being typically located at frequencies much larger than the LIGO/Virgo/KAGRA range, if its amplitude is large enough, there is hope that in the future this might be observed, revealing extremely valuable information about the early universe. Furthermore, oscillon-like objects might be relevant to address a number of phenomenologically interesting questions, such as dark matter~\cite{hep-ph/9303313, 1906.06352, 1909.11665, 1909.10805, 1912.07064, Boskovic:2018rub, Ferreira:2017pth, Annulli:2020lyc, Annulli:2020ilw, Khmelnitsky:2013lxt} and baryogenesis~\cite{1408.1811}.

Oscillons are meta-stable solutions for real scalar field theories~\cite{Bogolyubsky:1976yu, hep-ph/9308279, hep-ph/9503217, hep-ph/0209358}. They appear as localised lumps of energy inside which the field oscillates around the minimum of the scalar potential that defines the model. They are meta-stable, i.e. their lifetime is much longer than the oscillation timescale, which is of order $\mathcal{O}(1/m)$, where $m$ is the mass of the scalar field. Oscillon-like objects are known with different names, according to the actual interaction that makes them quasi-stable, see~\cite{1806.04690} for a recent review and a classification of the known compact objects allowed in a scalar field theory. What is traditionally referred to as an oscillon\footnote{Although oscillons were originally named \textit{pulsons}~\cite{Bogolyubsky:1976yu} and then \textit{breathers}~\cite{Segur:1987mg}.} is a compact object whose stability is ensured by the attractive scalar field self-interactions. In order for the model to feature attractive interactions, the scalar potential needs to be \textit{shallower than quadratic}. However, there are other mechanisms to ensure the stability of the oscillon: for instance if gravity plays a role, then the compact object is known as \textit{oscillaton} or \textit{real scalar star}~\cite{gr-qc/0301105, gr-qc/0104093, UrenaLopez:2002gx, UrenaLopez:2012zz}. The are also other variants of the same idea that include two fields, such as \textit{Q-balls}~\cite{Coleman:1985ki} and \textit{boson stars}~\cite{Jetzer:1991jr}, and exploit a global $U(1)$ symmetry to ensure their stability. Oscillon-like objects can also be formed in several non-standard scenarios, see for instance~\cite{hep-th/0610267, 0808.0514, 1210.7568, Antusch:2015ziz, Amin:2013ika, 1809.07724}. In this paper we will generically refer to pseudo-soliton solutions of a real scalar field theory as \textit{oscillons}, regardless of whether gravity plays a role in the meta-stability or not.

Since their discovery, there has been much interest in studying the lifetime and stability of oscillons in different regions of the parameter space spanned by their total mass, and the form of the scalar field potential. Oscillons decay classically by emitting scalar waves, and several works have analytically estimated and numerically computed the oscillon lifetime for various models~\cite{Segur:1987mg, Piette:1997hf, 0812.1919, 0903.0953, 0910.5922, 1003.3459, 1201.1934, 1210.2227, 1612.07750, 1901.06130, 1906.04070, Antusch:2019qrr, 1908.11103, 1911.03340, 2003.10899, Zhang:2020bec}. 

\begin{figure}\centering
\includegraphics[scale=0.45]{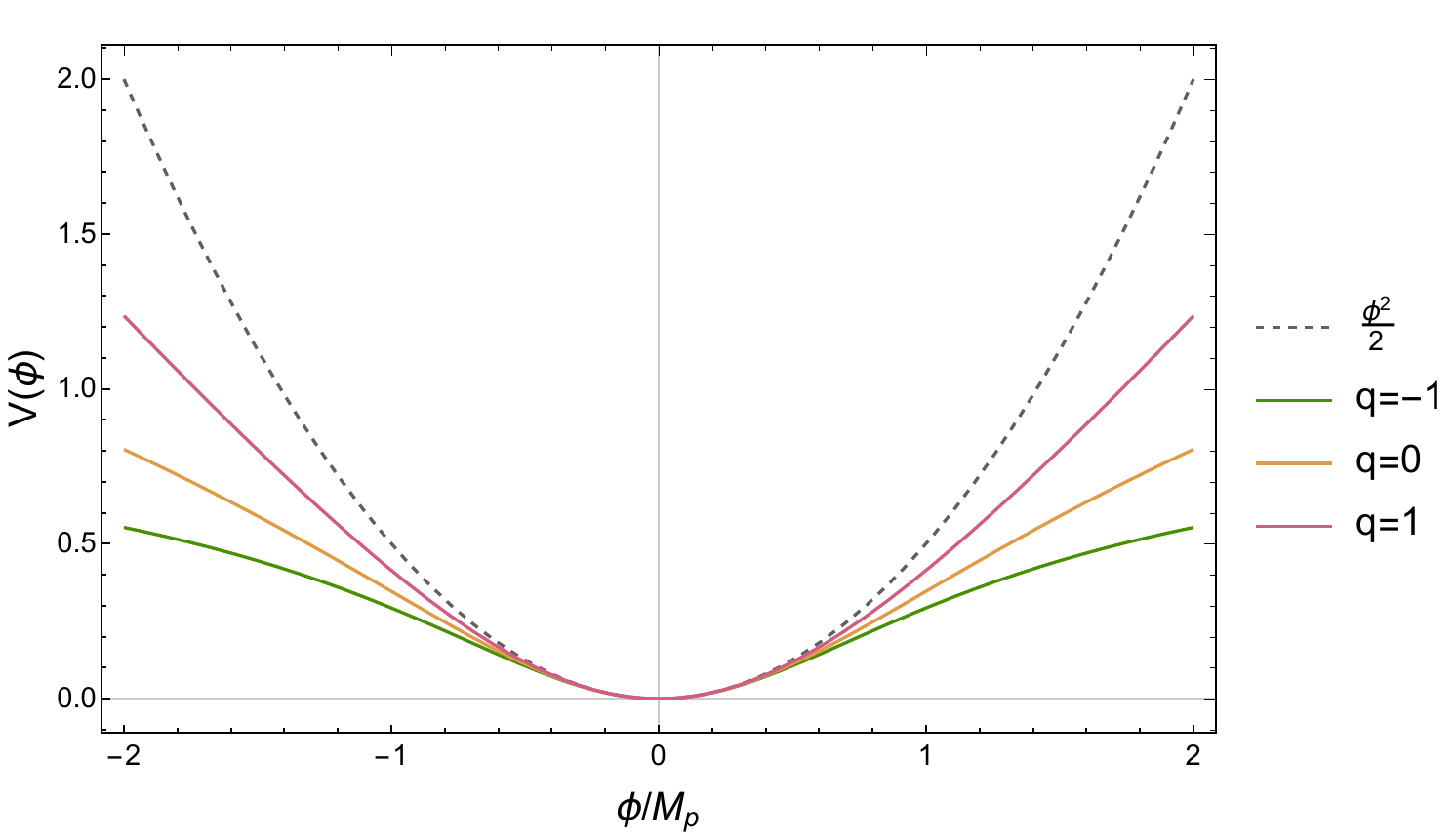}
\includegraphics[scale=0.45]{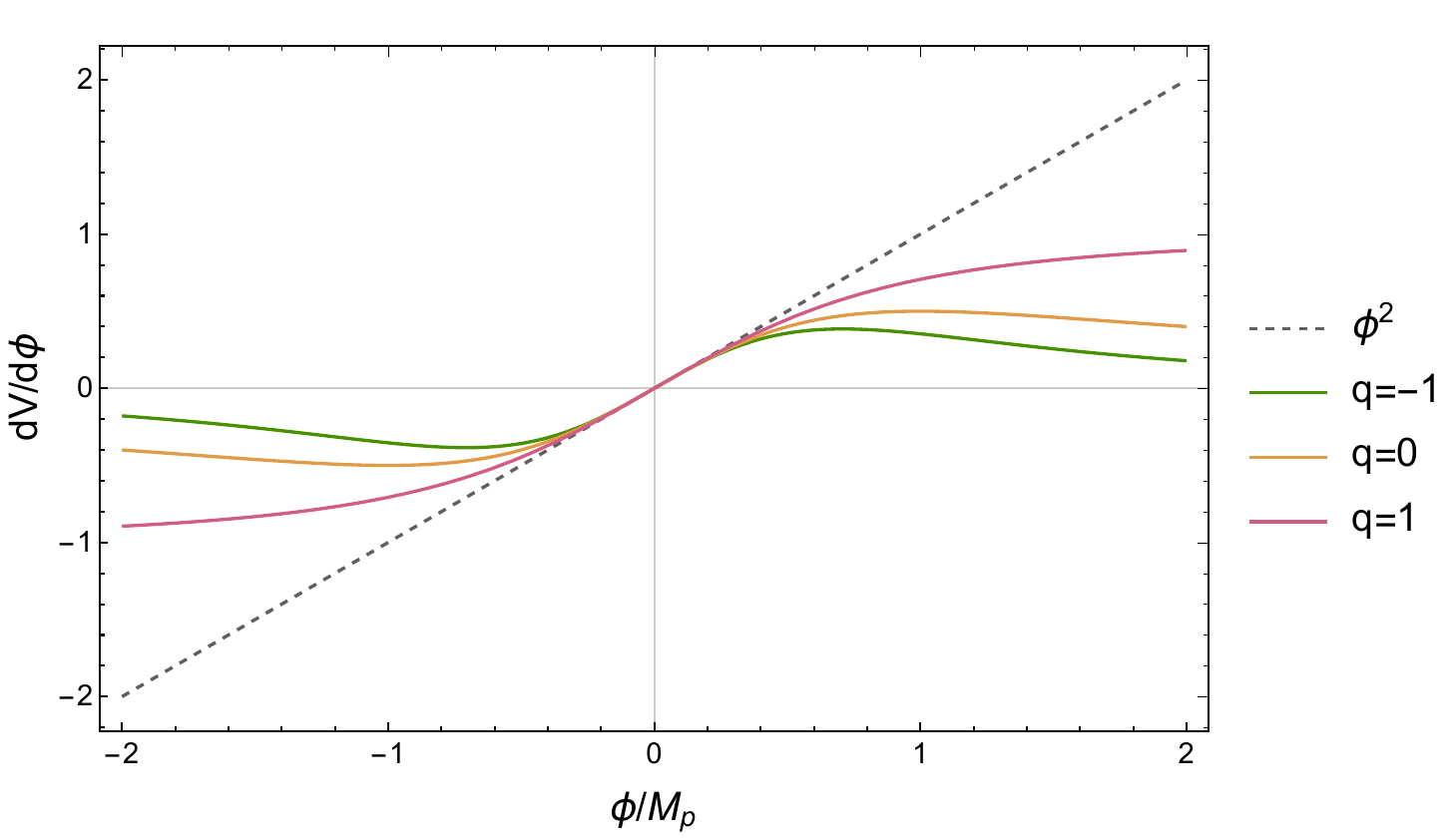}
\caption{We plot the potential $V(\phi)$ (top) and gradient $dV/d\phi$ (bottom) of Eq.~\eqref{eq:Potentials} for $q = -1, 0, 1$ and $\Lambda/M_p=1$, compared to the quadratic potential.
\label{fig:Potentials}}
\end{figure}

Our work continues these investigations: while previous studies of oscillon stability typically neglect gravitational effects, these become important in some regions of the parameter space, as already shown in~\cite{1807.09795, 1609.04724, 1708.01344, 1906.09346, 1907.10613, 1912.09658}. In this paper we will be primarily interested in strong gravity effects, which are relevant when the compactness of the oscillon $M/(M_p^2 R)$ (where $M$ is its total mass while $R$ is its radius) is of order $\sim 0.1$ and $\phi/M_p \sim 1$. For many problems with lower compactness (that in scalar field systems typically correspond to small displacements of the field $\phi/M_p \ll 1$), the weak gravity regime is sufficient to describe the dynamics. It has been shown that Newtonian gravity effects may stabilise low-density pseudo-solitons, see for instance~\cite{Hu:2000ke, Marsh:2013ywa, Schive:2014dra, Schive:2014hza, Hui:2016ltb, Marsh:2015wka} for interesting applications to the problem of dark matter.

In this paper we will explore a class of potentials of the form
\begin{equation}
V(\phi)=\frac{m^2 \Lambda^2}{q}\left[ \left(1+\frac{\phi^2}{\Lambda^2}\right)^{\frac{q}{2}}-1\right] \,,
\label{eq:Potentials}
\end{equation}
that are inspired by axion-monodromy~\cite{0803.3085, McAllister:2014mpa}. These potentials are quadratic around the minimum at $\phi=0$, $m$ being the mass of the scalar field. We will consider three cases which we identify by the value for the parameter $q=-1,0,1$, leading to three different characteristic behaviours of the potential. Note that what we refer to as the $q=0$ case is the limit of the potential for $q \rightarrow 0$ and thus has the logarithmic form
\begin{equation}
V(\phi)=\frac{m^2 \Lambda^2}{2} \ln \left(1+\frac{ \phi^2}{\Lambda^2}\right) \,.
\end{equation}
This range of values for the parameter $q$ ensures that the potential features a region that is shallower than quadratic at $\phi \gtrsim \Lambda$. The scale $\Lambda$ sets the characteristic energy scale of the potential, that determines the transition between the quadratic region and the shallower than quadratic one, see top panel of Fig.~\ref{fig:Potentials} for a visual comparison in the case $\Lambda/M_p = 1$.

References~\cite{1106.3335, 1906.06352, Zhang:2020bec} note that for the values of the parameter $q$ considered here, the lifetime of the corresponding oscillons is $\tau \gtrsim 6 \times 10^8 \, m^{-1}$ for $q=-1$, $\tau \gtrsim 3 \times 10^7 \, m^{-1}$ for $q=0$ and $\tau \gtrsim 10^8 \, m^{-1}$ for $q = 1$. The analyses performed in~\cite{1106.3335, 1906.06352, Zhang:2020bec} are independent of the actual value of the parameter $\Lambda$: what matters in the absence of gravity - beyond the value of the mass of the field $m$, that sets the characteristic timescale of the system - is the ratio $\phi/\Lambda$. This ratio has to be $\phi/\Lambda \gg 1$ inside the oscillon, in order for the particles to feel the attractive interactions: for $\phi/\Lambda \ll 1$ the potential is well approximated with a quadratic function. However, similarly to what was already observed in~\cite{1906.09346} for different classes of models, we expect that in the class of potentials described by Eq.~\eqref{eq:Potentials}, strong gravity effects become important as soon as $\Lambda$ becomes of order $\mathcal{O}(0.1-1) M_p$, and the objects may undergo collapse to black holes. Therefore in regions where $\phi \sim \Lambda \sim M_p$ both gravity and self-interactions will play a role, resulting in potentially interesting dynamics.

In this article we numerically study the dynamics of oscillons in this regime. Since in this regime we expect the stability of the oscillon configuration to be determined mainly by gravity instead of the attractive scalar field self-interaction, we proceed as follows
\begin{itemize}
\item Starting with known free field solutions for the profile of $\phi$ (which would be supported by gravity, and stable in the absence of self-interactions), we investigate how turning on self-interactions affects their stability for the three classes of potential parametrised by $q$, and three different values of $\Lambda$.
\item For comparison, we also investigate whether the free-field solutions will remain stable in the absence of gravity, by setting $G=0$ in our simulations. This tests the importance of the gravitational interaction in stabilizing the localized perturbation in the presence of self-interactions, starting from a free-field solution.
\end{itemize}
It should be noted that the free-field solutions that we use as initial conditions will not, in general, have the same profile as oscillons of the same mass supported by self-interactions alone, or equally for the gravity + self-interactions case. Therefore even if we see a collapse or dispersal, we cannot unambiguously state that no solution exists for this oscillon mass in the relevant regime - we may simply be starting too far away from the correct profile to converge on it. However, it provides an indication that the regime is likely to be unstable. If on the other hand we do not find a prompt collapse or dispersal, we consider that to be a quasi-stable configuration. Our simulations are too short to follow the full lifetime of the oscillon in this case, but we continue them for long enough to see that they are settling into a relatively steady oscillating state, albeit one that is likely to be excited relative to the ground state configuration.

This paper is organised as follows: in Sec.~\ref{sec:NumericalSetup} we explain the numerical setup, including how we initialise oscillons and the numerical method employed for our study; in Sec.~\ref{sec:ModelsAndResults} we describe our findings for each model that we consider; in Sec.~\ref{sec:Conclusions} we briefly summarise and discuss the results of the paper.

Throughout this paper we use Planck units with $M_p = \sqrt{\hbar c/8\pi G}$. Our simulations use geometric units in which $G=c=1$ and $m/\hbar = 1$. Therefore lengths and times in our plots are expressed in terms of the scale $1/m$ - the scalar mass is effectively absorbed in the choice of units for length and time.

\section{Numerical Setup}
\label{sec:NumericalSetup}

\subsection{Numerical Methods}
\label{sec:NumericalMethod}

We use $\textsc{GRChombo}$, an open source numerical relativity code, to evolve initial profiles of the scalar field coupled minimally to gravity in full general relativity. This allows us to accurately study oscillon masses right up to and beyond the limits of black hole formation. 
Full details of the GRChombo code can be found at the website~\url{www.grchombo.org}, and also in~\cite{Clough:2015sqa}. 
Briefly, GRChombo evolves the Einstein equation with the BSSN/CCZ4 formulation 
and the method of lines, with 4th order finite difference stencils in space and 4th order Runge-Kutta time integration. To follow black hole formation a moving puncture gauge is employed. This is a dynamical gauge choice which maintains coordinate observers at approximately fixed positions relative to the centre of the domain,
avoiding the focusing in overdense regions which occurs for geodesic observers.

The 3+1D ADM decomposition of the metric reads
\begin{equation}
ds^2 = - \alpha^2 dt^2 + \gamma_{ij} (dx^i + \beta^i dt) (dx^j + \beta^j dt) \,, \label{eq-metric}
\end{equation}
where $\alpha$ is the lapse and $\gamma_{ij}$ is the three-metric on the equal time hypersurfaces. In the BSSN formalism the induced metric is further decomposed as 
\begin{equation}
\gamma_{ij}=\frac{1}{\chi}\,\tilde\gamma_{ij}\,, \quad \det\tilde\gamma_{ij}=1\,, \quad \chi = \left(\det\gamma_{ij}\right)^{-\frac{1}{3}}  \,.
\label{eq-chi}
\end{equation}
We will frequently present the evolution of $\chi$ as representative of the evolution of the spacetime, with a value of $\chi$ which falls to zero being indicative of black hole formation (which is then confirmed by looking for trapped surfaces).

The simulations we present have spherical symmetry, but GRChombo uses a Cartesian grid. In general we use a coarsest domain of $128^3$ points to represent a region of $L = 128 ~ [1/m]$ with the oscillon at the centre. Using the octant symmetry of the problem in Cartesian coordinates (coming from the spherical symmetry of the setup) we only actually need to evolve 1/8th of the volume ($64^3$ points), with reflective boundary conditions in the negative x, y and z directions which take into account the appropriate parity of the gravitational components in each direction. We enforce four additional 2:1 refinement levels on top of this, which is sufficient to resolve non collapsing cases. However, we add additional refinement dynamically in order to keep the fields well resolved, which is activated during any collapse to a black hole. This refinement is driven by the gradients in $\chi$. We present the some code validation checks in Appendix \ref{sec-convergence}.

\subsection{Initial Conditions}
\label{sec:InitialConditions}

Meta-stable solutions for the real free scalar field case can be numerically obtained by solving the Einstein-Klein-Gordon (EKG) equations for a scalar field profile with the ansatz of a harmonic time dependence, see for example~\cite{UrenaLopez:2002gx, 1609.04724}. In that case, a shooting method is used to construct a one-parameter family of meta-stable solutions, characterised by the maximum amplitude of the scalar field at the centre of the oscillon, denoted by $\phi_0/M_p$, or equivalently by their total mass $M$, usually expressed in units $[M_p^2/m]$ (see table Tab.~\ref{tab:1to1Relation} for the conversion). These meta-stable solutions can be further characterised as \textit{stable} under small perturbations, if $\phi_0/M_p < 0.48$, and unstable under small perturbations, if $\phi_0/M_p > 0.48$.  

As noted in the Introduction, we will use the field profiles of solutions that are stable under small perturbations in the free-field case as initial conditions for our simulations, as described below. (Note that we parameterise these in terms of the mass $M$ since this is still a relevant physical quantity when the potential includes self-interactions, whereas the peak amplitude of the oscillation $\phi_0$ is likely to change beyond the free-field case.)

\begin{table}[h!]
\centering
\begin{tabular}{|c|c|}
\hline
$\phi_0 ~ [M_p]$ & $M ~ [M_p^2/m]$ \\ \hline
$0.10$ & $2.07$ \\ \hline
$0.20$ & $2.63$ \\ \hline
$0.30$ & $2.90$ \\ \hline
$0.40$ & $3.01$ \\ \hline
$0.48$ & $3.04$ \\ \hline
\end{tabular}
\caption{One-to-one relation between the value of the maximum central amplitude of the free-field oscillon and its total mass. We characterise our initial conditions in terms of the total mass of the oscillon, as the resulting amplitude may differ under evolution in the non linear potential. \label{tab:1to1Relation}}
\end{table}

\begin{figure}[h!]\centering
\includegraphics[scale=0.8]{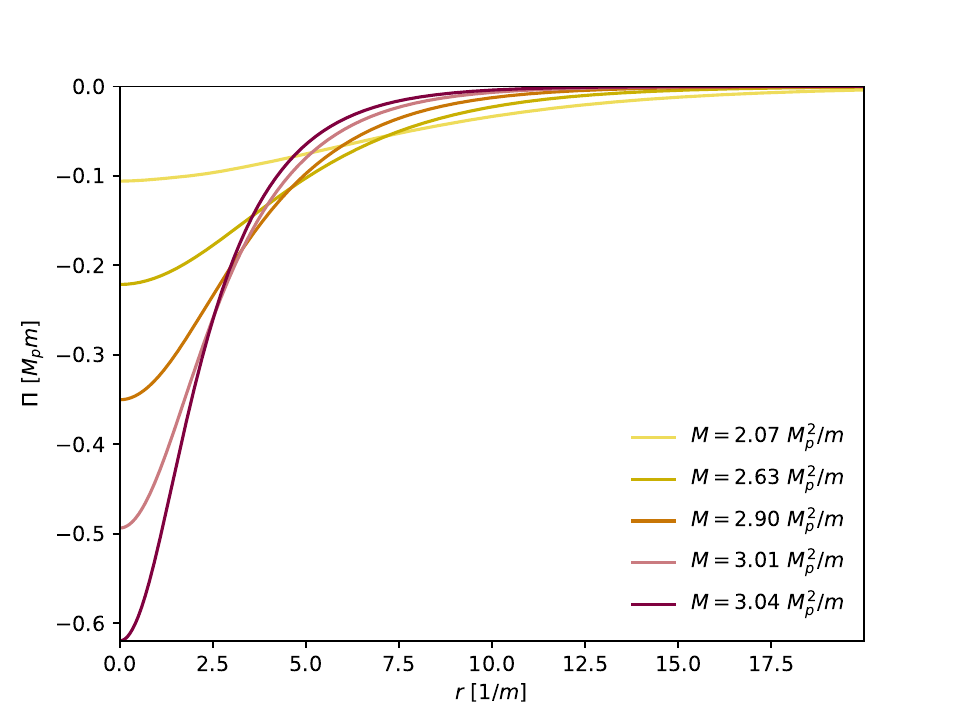}
\caption{Spatial profiles of the initial condition for the conjugate momentum $\Pi = \frac{1}{\alpha}(\dot\phi - \beta^i \partial_i \phi)$ of the field, for different total masses of the oscillon $M$. \label{fig:pi}}
\end{figure}

In models with a more complicated scalar potential it is not straightforward to solve the EKG equations to find meta-stable solutions and their stability limits, hence in this paper we use the free-field solutions as approximate guesses, and test whether they evolve into stable solutions or disperse/collapse once self-interactions are included. When the potential is close to the simple massive case (cases for which $\Lambda \gg M_p$), the profiles will remain constant and the same stability bounds will be obtained, but for the strongly attractive interactions studied here $\Lambda \lesssim M_p$ and $\phi \gtrsim \Lambda$, we expect potentially significant deviations in both the profile and the stability.
To be able to use the free-field solutions with an arbitrary potential, we pick the point of time symmetry in which
\begin{equation}
\phi(r) = 0 \,, \qquad \dot\phi(r) \neq 0 \,.
\end{equation}
In this way, the momentum constraint\footnote{$K_{ij}$ and $K$ are the extrinsic curvature and its trace respectively, while $D_j$ is the covariant derivative with respect to $\gamma_{ij}$.}
\begin{equation}
\mathcal{M}_i = D_i K - D^j K_{ij} - 8 \pi S_i = 0 \,,
\end{equation}
is trivially satisfied because the gradients of the field vanish everywhere, thus the momentum density $S_i = \gamma_{ia} \gamma_{jb} T^{ab} \sim \dot\phi(r) \partial_i \phi = 0$ ($T^{ab}$ is the stress-energy tensor). At the same time, the only additional terms in the Hamiltonian constraint\footnote{Where $\rho$ is the energy density and ${}^{(3)} R$ is the $3$-curvature.}
\begin{equation}
\mathcal{H} = {}^{(3)} R + K^2 + K_{ij} K^{ij} - 16 \pi \rho = 0 \,, \label{eqn-Ham}
\end{equation}
with respect to the free-field case are the potential terms beyond $\frac{1}{2} m^2 \phi^2$, but those vanish because $\phi(r) = 0$ everywhere. Hence, given the solutions for the free-field case (which we obtained using a shooting method) the Hamiltonian constraint is satisfied at this point of time symmetry, and we can use the free-field solution as the initial condition for our simulations without violating the constraints. This leads to a non trivial profile for the field time derivative $\Pi \sim \dot \phi$, and the lapse $\alpha$ and conformal factor of the metric $\chi$, as defined in equations (\ref{eq-metric}) and (\ref{eq-chi}) above. The initial solutions for $\Pi$ are illustrated in Fig.~\ref{fig:pi}, from which we see that more massive solutions are more compact with a higher central amplitude in the field and a smaller spatial extent. The spatial metric is chosen initially to be conformally flat, so the numerically obtained solutions in areal polar coordinates are transformed into conformally flat ones and then interpolated onto the numerical grid using a first order scheme. 

As noted above, despite these solutions being meta-stable in the free-field case, they do not necessarily maintain this property once interactions are taken into account, and so they may be poor initial guesses for stable configurations. However, it was observed in~\cite{1906.09346} that if a meta-stable solution exists for the models considered, and if it is not too far from the free-field meta-stable solution, then the initial conditions that we use here tend to be dynamically driven towards the new meta-stable solutions. We effectively start with an ``excited'' state of the true solution, and the additional excitations are radiated over a transient period.
This is the behaviour that we observe in most examples in this paper, see Sec.~\ref{sec:ModelsAndResults}. However, we also find some unexpected instabilities, which may be at least partly due to the initial conditions being too far from the meta-stable solutions to converge on them. This is discussed further below.

\section{Models and Results}
\label{sec:ModelsAndResults}

As detailed in Sec.~\ref{sec:Introduction}, we study the class of potentials described in Eq.~\eqref{eq:Potentials}. For each of the 3 values of the parameter $q$, we set the initial profile of the oscillon to the free-field, gravitationally supported solution, according to the procedure described in Sec.~\ref{sec:InitialConditions}, with 5 different values of the total mass $M$ (see Tab.~\ref{tab:1to1Relation}). For each of these values, we simulate the subsequent dynamics once self-interactions are turned on, firstly including gravity and then removing gravitational support by setting the gravitational constant $G = 0$, and fixing the background as Minkowski space. We explore the differences in the evolution, noting whether the result is a collapse, dispersion, or meta-stability. 
We also explore 3 different energy scales for the self-interaction term $\Lambda/M_p = 0.1,~ 0.3,~ 1.0$. Above the upper limit of this range we find that the behaviour and solutions tend to that of the free-field case. Below the lower limit the impact of the potential on the gravitational field becomes negligible and the behaviour approaches that of a massless scalar field.
For each choice of the parameter $q$ we display the resulting dynamics in a table, denoting by MS the cases that feature a meta-stable solution at the end of the simulation, by D the cases that disperse during the simulation and by C the cases that lead to a collapse, namely to black hole formation. Furthermore, we show the time evolution of the scalar field $\phi_0$ and of the conformal factor $\chi$ at the centre of the oscillon for several interesting cases.

\subsection{Case $q = -1$}
\label{sec:qminus1}

The potential takes the form
\begin{equation}
V(\phi) = m^2 \Lambda^2 \left[1-\frac{1}{\sqrt{1+\frac{ \phi^2}{\Lambda^2}}} \right] \,.
\end{equation}
The dynamics for the case $q = -1$ are summarised in Tab.~\ref{tab:qminus1}, with the full numerical results given in Appendix \ref{app:qminus1}.

\begin{table}[]
\centering
\begin{tabular}{||c||c|c||c|c||c|c||}
\hline 
\multirow{2}{*}{$q=-1$}& \multicolumn{2}{c||}{$\Lambda/M_p=0.1$} & \multicolumn{2}{c||}  {$\Lambda/M_p=0.3$} & \multicolumn{2}{c||}{$\Lambda/M_p=1.0$} \\ \cline{2-7} 
& $G=0$ & $G=1$ & $G=0$ & $G=1$ & $G=0$ & $G=1$ \\ \hline\hline
$M = 2.07 \, M_p^2/m$&MS &MS&MS&MS&D&MS\\ \hline
$M = 2.63 \, M_p^2/m$&MS & MS &MS&MS&D&C\\ \hline
$M = 2.90 \, M_p^2/m$&  MS& MS &MS&MS&D&C\\ \hline
$M = 3.01 \, M_p^2/m$& MS& MS &MS&MS&D&C\\ \hline
$M = 3.03 \, M_p^2/m$ &  MS& MS &MS&MS &D&C \\ \hline
\end{tabular}
\caption{Summary of the stability of the initial configuration in the case $q=-1$.\label{tab:qminus1}}
\end{table}

\begin{figure}[h!]\centering
\begin{minipage}{\textwidth}
\includegraphics[scale=0.95]{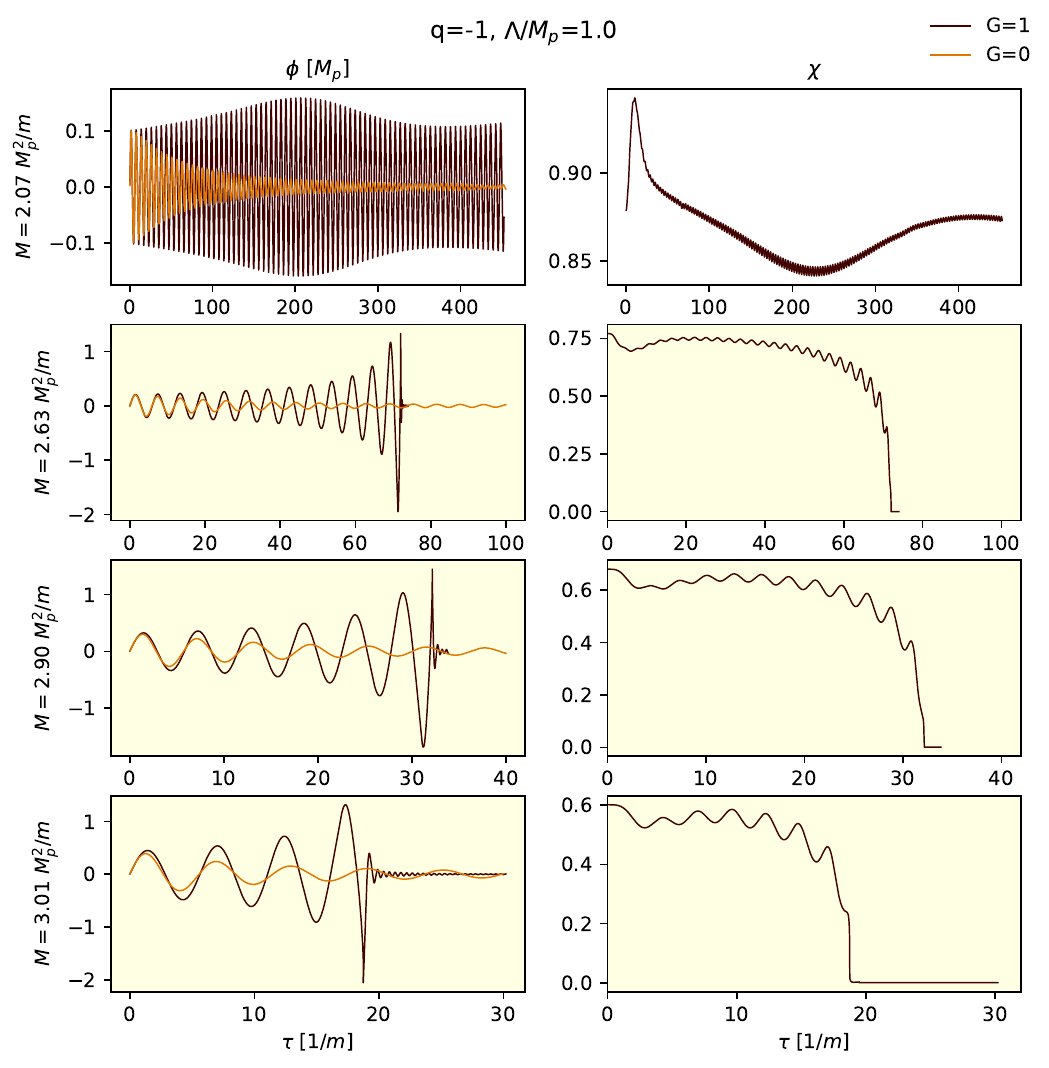}
\caption[caption]{Field dynamics in the case $q=-1$, $\Lambda/M_p = 1$. The left column refers to the field at the centre of the oscillon in Planck units $\phi/M_p$, while the right column of plots refers to the conformal factor $\chi$ at the centre of the oscillon. We find that only the first value of $M=2.07 \, M_p$ gives a meta-stable solution. The remaining cases collapse to black holes due to the effect of the additional attractive self-interactions. \label{fig:qminus1} \footnotemark}
 \end{minipage}
\end{figure}
\footnotetext{Throughout the paper we use brown lines for cases which include gravity, and orange for
non-gravity cases. We also use a shaded background to highlight those that collapse to black holes in the gravity case.}
Fig.~\ref{fig:qminus1} shows the central field and conformal factor profiles in the case $\Lambda/M_p = 1.0$, $M \geq 2.07 \, M_p^2/m$. Recall that the conformal factor dropping to zero signals the formation of a horizon, thus we see that the three higher mass cases all result in collapse to a black hole, with only the first being meta-stable. When gravitational support is removed, all these cases disperse, indicating that the self-interaction alone is not sufficient to support the configuration, which is expected since in this case $\phi < \Lambda$.

This is therefore a clear case where taking into account the effects of both gravity and self-interactions might drastically change the resulting dynamics and stability of oscillons in the model. We see that adding the attractive interactions, whilst stabilising oscillons in smaller amplitude cases, results in lower stability than a purely massive potential in the strong gravity regime. This can be understood by considering the potential gradient in the bottom panel of Fig.~\ref{fig:Potentials} and the dynamics of the central value of the field, where
\begin{equation}
    \ddot \phi \sim \nabla^2 \phi - \frac{dV}{d\phi}.
\end{equation}
We see that a flatter potential results in a {\it smaller} restoring force, and thus large excursions of the field on the potential have a tendency to destabilise the free-field configuration, leading to its collapse. There is, in effect, too much attraction when both gravity and the self-interaction are included.

\begin{figure}[h!]\centering
\includegraphics[scale=0.96]{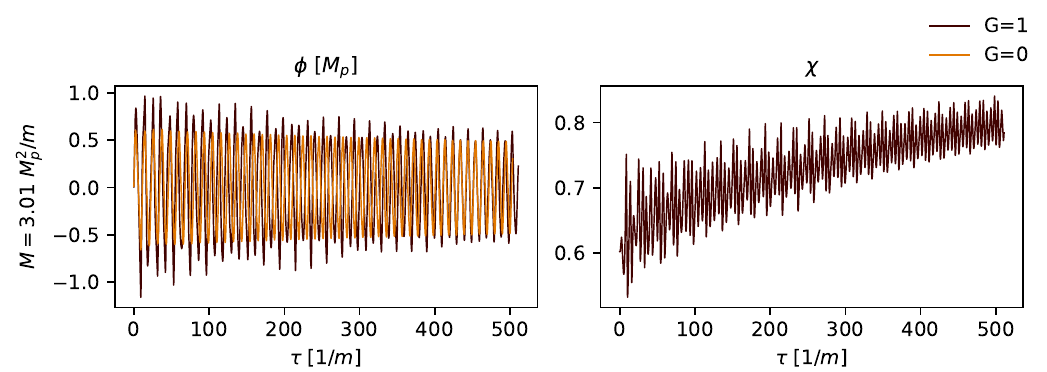}
\caption{Field dynamics in the case $q=-1$, $\Lambda/M_p = 0.1$. The left column refers to the field at the centre of the configuration $\phi$, while the right column refers to the conformal factor $\chi$ at the same point. For smaller values of $\Lambda$ we find a meta stable oscillon both with and without gravity for all the values that we tested, once self-interactions are added. \label{fig:qminus01}}
\end{figure}

For the cases with $\Lambda/M_p = 0.1$ and $\Lambda/M_p = 0.3$ all values appear to be meta-stable. In these cases the value of $V(\phi)$ is smaller for the same amplitude of the field and so the gravitational effects are weaker - the combination of attractive self-interactions and gravitational attraction seems to find a stable balance. The field radiates energy during the first oscillations and settles into an excited but meta-stable state. This is true both in the cases that include gravity and in the cases with $G = 0$ where gravity is neglected. The main difference between these two regimes is in the amplitude of the oscillations - as shown in Fig.~\ref{fig:qminus01} gravity makes them somewhat larger. As discussed for instance in~\cite{Antusch:2017vga}, as soon as the assumption about spherical symmetry is relaxed, a larger amplitude of the field inside the oscillon might lead to a larger GW production in a scenario in which these oscillons are produced after preheating. 

\subsection{Case $q = 0$}
\label{sec:q0}

In this case the potential features a logarithmic dependence on the scalar field
\begin{equation}
V(\phi)=\frac{m^2 \Lambda^2}{2} \ln \left(1+\frac{ \phi^2}{\Lambda^2}\right) \,.
\end{equation}
The behaviour is summarised in Tab.~\ref{tab:q0}, with the full numerical results given in Appendix \ref{app:q0}, where we see that the dynamics is more complicated in this case. We observe three different behaviours for the three choices of $\Lambda/M_p$

\begin{table}[h!]
\centering
\begin{tabular}{||c||c|c||c|c||c|c||}
\hline 
\multirow{2}{*}{$q=0$}& \multicolumn{2}{c||}{$\Lambda/M_p=0.1$} & \multicolumn{2}{c||}  {$\Lambda/M_p=0.3$} & \multicolumn{2}{c||}{$\Lambda/M_p=1$} \\ \cline{2-7} 
 &$G=0$&$G=1$&$G=0$&$G=1$&$G=0$&$G=1$\\ \hline\hline
 $M = 2.07 \, M_p^2/m$&MS & MS&D&MS&D&MS\\ \hline
 $M = 2.63 \, M_p^2/m$& MS&  C&D&C&D&C\\ \hline
 $M = 2.90 \, M_p^2/m$&  MS&  C&D&C&D&C\\ \hline
 $M = 3.01 \, M_p^2/m$& MS &  MS&D&MS&D&C\\ \hline
  $M = 3.03 \, M_p^2/m$ &  MS&MS  &D&MS &D&C \\ \hline
\end{tabular}
\caption{Summary of the stability of the initial configuration in the case $q=0$.\label{tab:q0}}
\end{table}

\begin{itemize}
\item For $\Lambda/M_p = 1$ we observe a behaviour similar to the case $q=-1$ described in Sec.~\ref{sec:qminus1}, where there is a prompt collapse to a black hole above $M \geq 2.07 \, M_p^2/m$, whilst the initial profile always disperses if the effects of gravity are neglected, see Fig.~\ref{fig:q0Lambda1}.

\begin{figure}[h!]\centering
\includegraphics[scale=0.93]{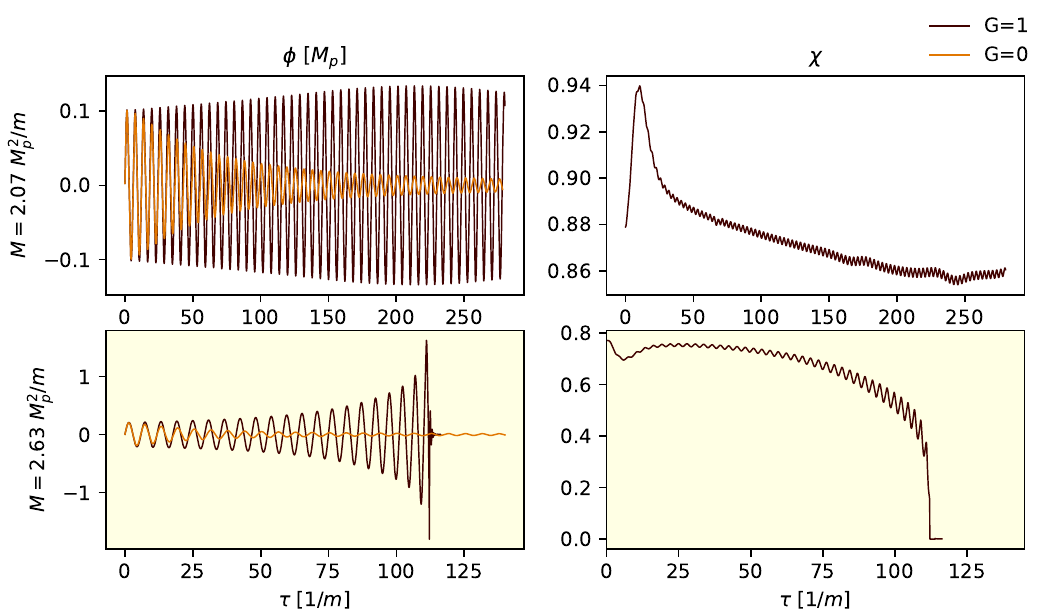}
\caption{As in Fig. \ref{fig:qminus1}, but for the case $q=0$, $\Lambda/M_p=1$. Again we see that there is a transition between stability and collapse above $M = 2.07\, M_p^2/m$, meaning that the configuration is less stable than the purely massive case. We also plot the cases without gravity, where we see that the field disperses - self-interactions are too weak to support the configuration alone where $\phi < \Lambda$. (Note that we cut the plots at the point where we no longer trust the dispersed cases due to large boundary reflections, but run the $G=1$ case longer to confirm stability for $M = 2.07 M_p^2 / m$.)}
\label{fig:q0Lambda1}
\end{figure}

\item At the other end of the scale, for $\Lambda/M_p = 0.1$, while neglecting the effects of gravity we always get meta-stable solutions (as might be expected since here $\phi > \Lambda$), the cases with both gravity and self-interactions can lead to black hole formation, see Fig.~\ref{fig:q0}. 
The fact that for intermediate values of the total initial mass ($2.63 \, M_p^2/m \leq M \leq 2.90 \, M_p^2/m$) we have black hole formation, while for greater values of $M$ ($M \geq 3.01 \, M_p^2/m$) we have again meta-stable oscillon states, is a rather surprising behaviour. We ascribe this peculiar feature to the sub-optimal initial conditions that we are using. As explained in Sec.~\ref{sec:InitialConditions}, using the free-field solutions as initial conditions does not guarantee that the perturbation is close enough to the actual meta-stable state (once interactions are taken into account) so that the latter is dynamically reached by the system. It might well be that, for some combination of the parameters that we are considering, the free-field solution is sufficiently different that we cannot reach it by dynamical evolution. This seems to be the main reason for the irregular behaviour that we are observing in these models, in which there is no clear boundary in parameter space between the collapse region and the meta-stable one. We note that a similar qualitative behaviour was observed in~\cite{1708.01344} for the case of axion stars with a cosine potential. The instability in this regime merits further investigation, and motivates the search for more accurate analytic initial conditions.

\begin{figure}[h!]\centering
\includegraphics[scale=0.93]{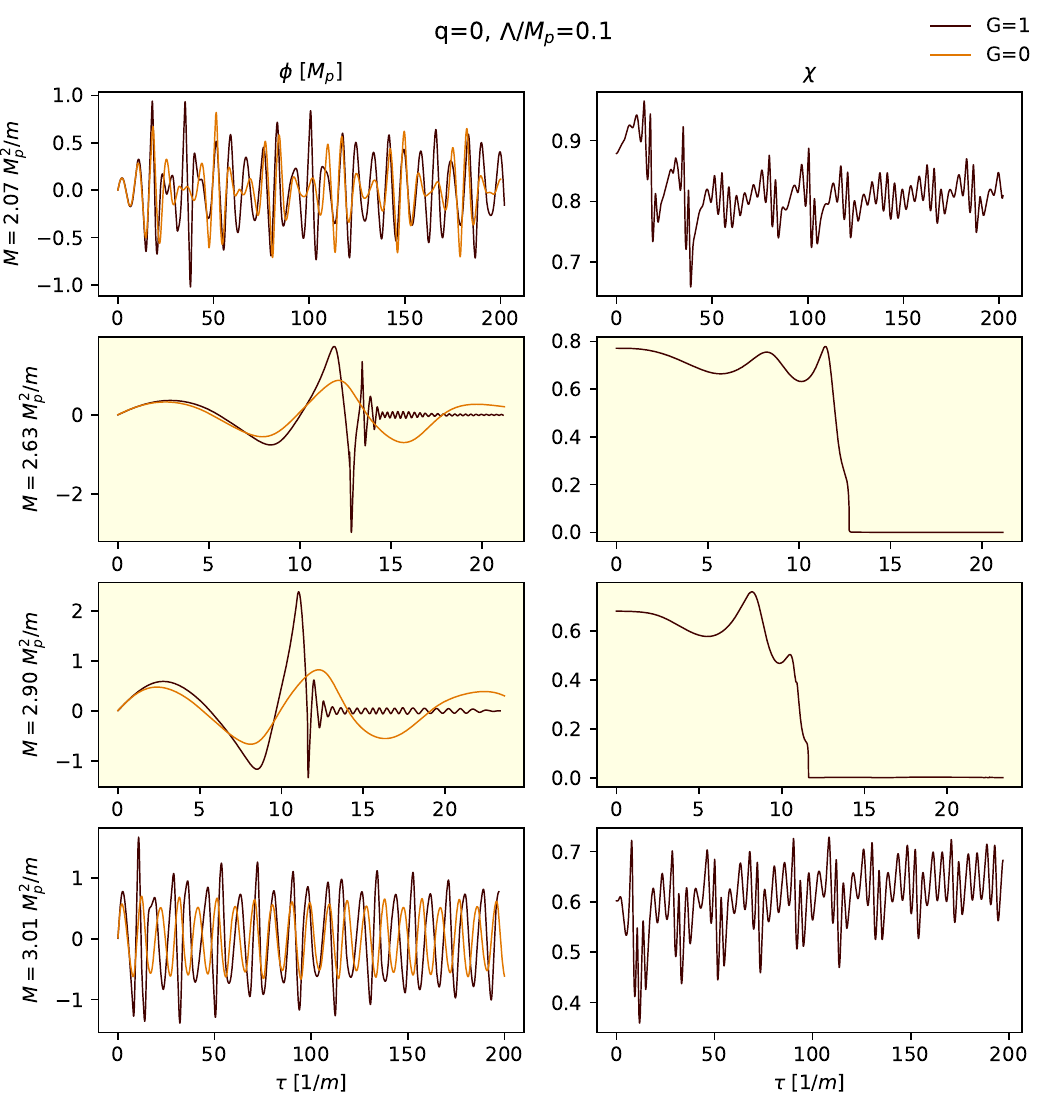}
\caption{Field dynamics in the case $q=0$, $\Lambda/M_p=0.1$. The left column refers to the field $\phi$ at the centre of the configuration, while the right column refers to the conformal factor $\chi$ at the centre of the oscillon. In this case we unexpectedly see a recovery in stability for the larger mass case, whilst smaller mass cases collapse to black holes. As discussed in the text, this may be due to the initial conditions being too far from stable solutions to converge on them dynamically. In the zero gravity case all cases are meta-stable.
\label{fig:q0}}
\end{figure}

\item For the intermediate case of $\Lambda/M_p = 0.3$ we see a behaviour similar to $\Lambda/M_p = 0.1$, except that if we only have self-interactions and no gravity, then the field profile quickly disperses. This is due to the fact that the field mainly probes the quadratic part of the potential, hence the attractive force is smaller and insufficient to keep the perturbation localised. 
\end{itemize}

\subsection{Case $q=1$}
\label{sec:q1}

The potential in this case takes the form
\begin{equation}
V(\phi)=m^2\Lambda^2\left[\sqrt{1+\frac{ \phi^2}{\Lambda^2}} -1\right] \,.
\end{equation}
The dynamics in this case are summarised in Tab.~\ref{tab:q1}, with the full numerical results given in Appendix \ref{app:q1}, where we see a behaviour similar to the $q=0$ case

\begin{table}[]
\centering
\begin{tabular}{||c||c|c||c|c||c|c||}
\hline 
\multirow{2}{*}{$q=1$}& \multicolumn{2}{c||}{$\Lambda/M_p=0.1$} & \multicolumn{2}{c||}  {$\Lambda/M_p=0.3$} & \multicolumn{2}{c||}{$\Lambda/M_p=1$} \\ \cline{2-7} 
 &$G=0$&$G=1$&$G=0$&$G=1$&$G=0$&$G=1$\\ \hline\hline
 $M = 2.07 \, M_p^2/m$& MS&MS &D&MS&D&MS\\ \hline
 $M = 2.63 \, M_p^2/m$&MS &  C&D&C&D&MS\\ \hline
 $M = 2.90 \, M_p^2/m$& MS &  C&D&C&D&C\\ \hline
 $M = 3.01 \, M_p^2/m$&  MS& C &D&C&D&C\\ \hline
 $M = 3.03 \, M_p^2/m$ &  MS& C &D&MS &D&C \\ \hline
\end{tabular}
\caption{Summary of the stability of the initial configuration in the case $q=1$.}
\label{tab:q1}
\end{table}

\begin{itemize}
\item For the case $\Lambda/M_p = 0.1$ we observe the behaviour shown in Fig.~\ref{fig:q1Lambda01}. In the absence of gravity the solution is always meta-stable - self-interactions are sufficient to support the profile, while with both self-interactions and gravity the profile collapses to a black hole for all cases above $M \geq 2.63 \, M_p^2/m$.

\begin{figure}[h!]\centering
\includegraphics[scale=0.93]{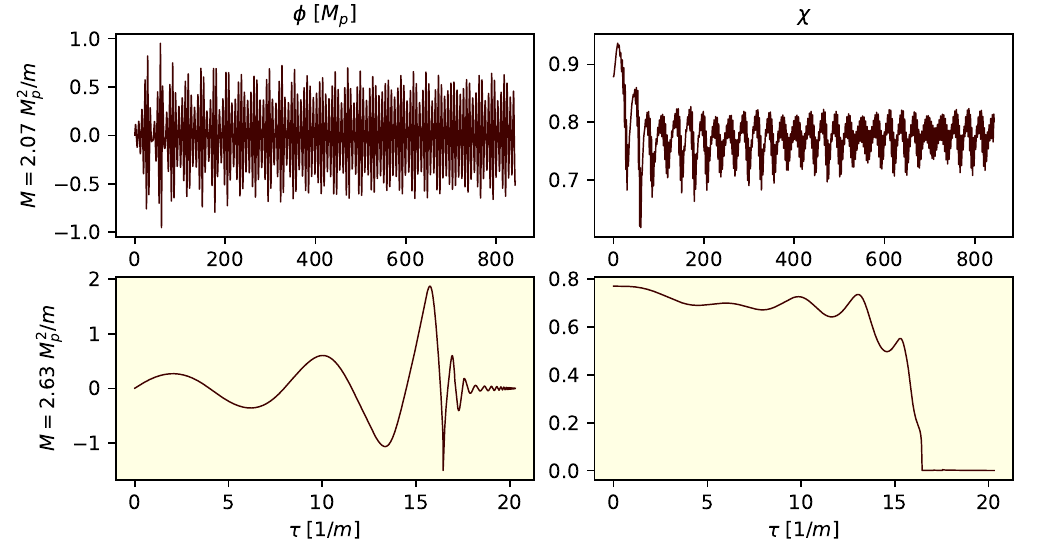}
\caption{Field dynamics in the case $q=1$, $\Lambda/M_p=0.1$. The left column refers to the field $\phi$ at the centre of the configuration, while the right column refers to the conformal factor $\chi$ at the same point. In this case we find that all cases except for the smallest mass case are unstable and collapse to black holes, whereas in the case without gravity they are meta-stable.  \label{fig:q1Lambda01}}
\end{figure}

\item For $\Lambda/M_p = 0.3$ the solutions always disperse with $G=0$, as the field mainly probes the quadratic part of the potential. On the other hand, the solution collapses to a black hole when $G=1$ and the initial total mass of the oscillon takes intermediate values $2.6 \, M_p^2/m \leq M \leq 3 \, M_p^2/m$. It appears to restabilise at the highest mass, which we attribute to the same reasons described in the $q=0$ case above.

\item For the case $\Lambda/M_p = 1$, we find collapse to a black hole for cases above mass $M \geq 2.6 \, M_p^2/m$. On the other hand, the profile always disperses if the effects of gravity are neglected, as illustrated in Fig.~\ref{fig:q1Lambda1}.

\begin{figure}[h!]\centering
\includegraphics[scale=0.93]{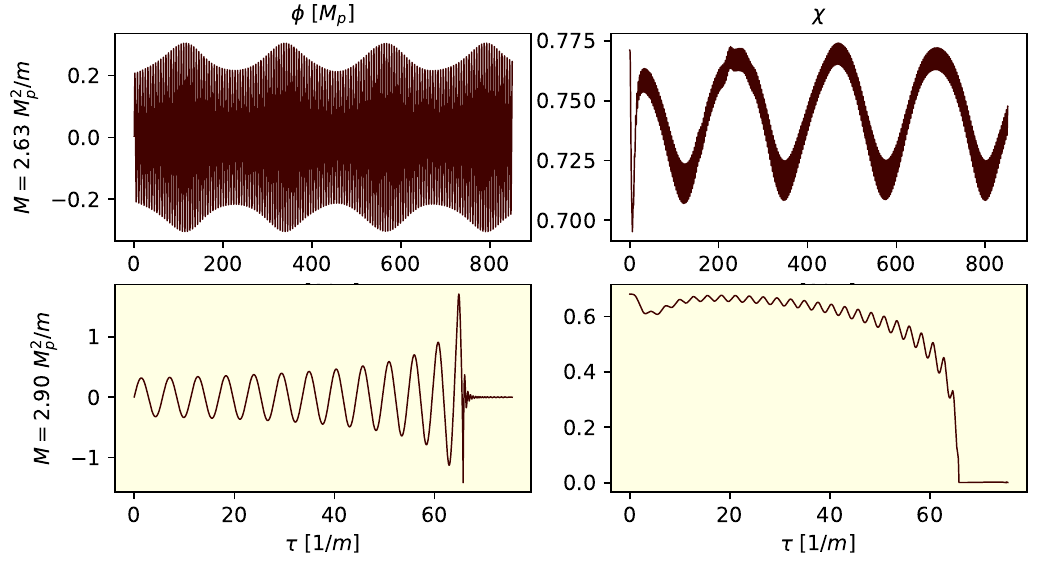}
\caption{Field dynamics in the case $q=1$, $\Lambda/M_p=1$. The left column of the plot refers to the field $\phi$ at the centre, while the right column refers to the conformal factor $\chi$ at the same point. We find stable oscillons at a higher mass than the $\Lambda/M_p=0.1$ case, but without gravity the field disperses.  \label{fig:q1Lambda1}}
\end{figure}

\end{itemize}

\section{Conclusions}
\label{sec:Conclusions}

\begin{figure}[h!]\centering
\includegraphics[scale=0.46]{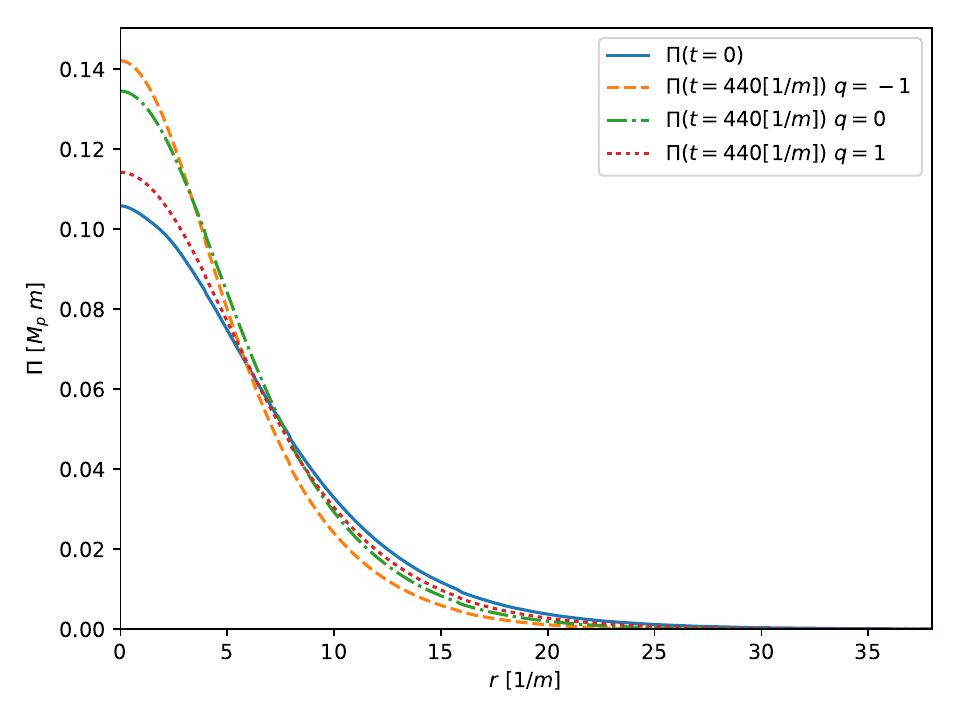}
\includegraphics[scale=0.46]{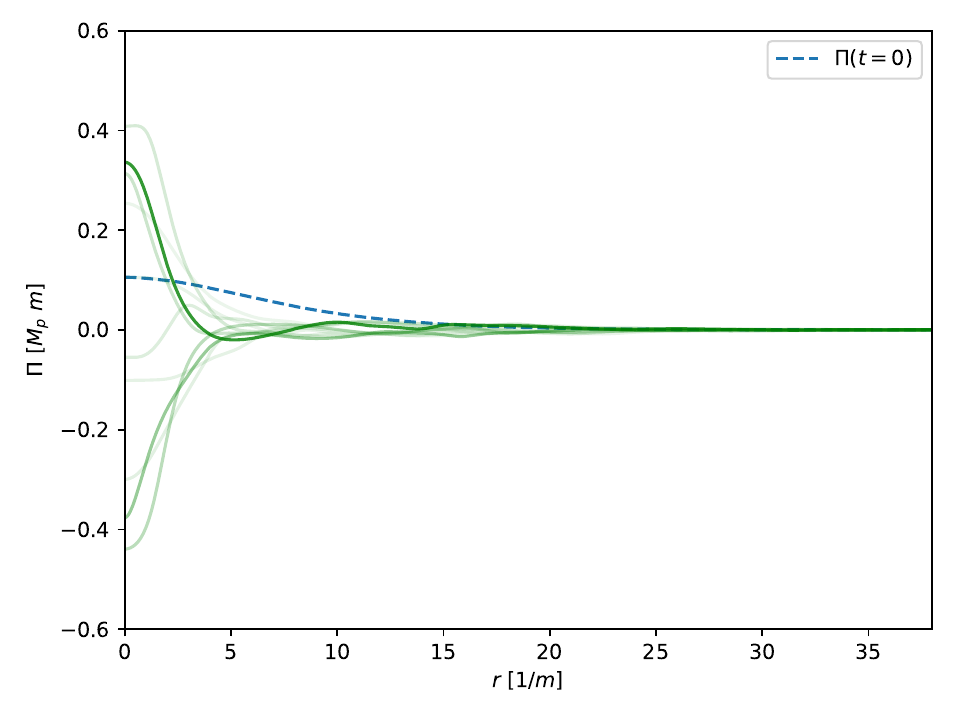}
\caption{Field profiles for cases with $M = 2.07 \, M_p^2/m$. The left panel shows the profiles in the case  $\Lambda=1 ~M_p$ for each of the three values of $q$ at a later time in the simulation.  The right panel shows a series of snapshots of the profile over time for the case $\Lambda=0.1 ~M_p$, $q=0$ with gravity. The attractive self-interactions result in a more peaked profile compared to the free-field case ($\nabla^2 \phi$ must be larger to balance the attractive self-interactions), which is most pronounced for the flattest potential - the $q=-1$ case. In the cases with $\Lambda=0.1 ~M_p$ where the amplitude of the oscillons strongly probes the flat part of the potential the profiles are somewhat ``messy''; they are initially in a superposition of excited states with different frequencies, which should decay away over time to leave the most stable ground state. \label{fig:fieldprofiles}}
\end{figure}

In this paper we expanded the studies in~\cite{1906.09346} to a wider class of potentials, tracking stability in different regions of the parameter space of the potential and configuration mass. We focused on axion-monodromy inspired models, whose potential takes the form in Eq.~\eqref{eq:Potentials}. We provide an analysis that is complementary to the one presented in~\cite{Zhang:2020bec}, covering regions of the parameter space where general relativity effects become important. We confirm the expectation that when the characteristic scale of the scalar potential is comparable to the Planck mass $\Lambda \approx M_p$ and the field probes the region $\phi \gtrsim \Lambda \sim M_p$, then the inclusion of general relativity effects becomes crucial in order to successfully study the dynamics of the system. We show in Fig. \ref{fig:fieldprofiles} field profiles of $\Pi$ to illustrate the dynamics. In particular, in this work, we have observed that
\begin{itemize}
    \item Spherically symmetric localized perturbations of scalar field, that cannot be supported by self-interactions alone, may be made stable by the addition of gravitational effects, see for instance the case $\Lambda/M_p = 1.0$, $M = 2.07 \, M_p^2/m$;
    \item Too much attraction can be a bad thing for stability - in some cases the interplay between gravitational attraction and self-interactions destabilises the perturbation, leading it to collapse to a black hole. This may happen in cases where the solution with $G=0$ gives rise to a meta-stable state (see for instance the case $q=0$, $\Lambda/M_p=0.1$, $M = 2.63 \, M_p^2/m$) and if the solution with $G=0$ disperses (see for instance the case $q=0$, $\Lambda/M_p = 0.3$, $M = 2.90 \, M_p^2/m$). As a general rule adding attractive self-interactions makes perturbations less stable than the free-field case;
    \item Using the initial conditions that we have chosen, we do not observe a sharp transition from the region in which oscillons are stable and the region in which they collapse to a black hole, which is similar to the findings in~\cite{Michel:2018nzt}. We ascribe this behaviour to the fact that the initial conditions that we using are the stable solutions for the free-field case (as described in Sec.~\ref{sec:InitialConditions}), which may be too far from any stable solutions in the modified potential to evolve dynamically towards them.
\end{itemize}
Given the last point, one could improve the study by updating the numerical shooting method used to find stable solutions to explicitly include the impact of self-interactions to some perturbative order, in order to start closer to the ``true'' solutions.
 
As the dynamics of oscillons studied in this paper is potentially relevant in the early Universe, it would also be interesting to extend the numerical analysis to the formation of these compact objects starting from vacuum fluctuations, or a more general power spectrum, after inflation. Such a study, including the effects of general relativity, would be relevant where significant overdensities develop which may collapse to black holes. Less challenging technically would be to generalise the analysis in this paper to multi-field and non-spherically symmetric scenarios, to see how these changes may affect the stability of compact objects. We plan to address these open questions in the near future.

\paragraph{Acknowledgments}
We acknowledge interesting discussions with Mustafa Amin and Edmund J. Copeland that inspired this work. We thank Francisco Pedro, Fernando Quevedo and Gian Paolo Vacca for collaboration in the previous project that laid the groundwork for this one.  ZN is supported by ICTP-Sandwich Training Educational Programme (STEP).  KC acknowledges funding from the European Research Council (ERC) under the European Unions Horizon 2020 research and innovation programme (grant agreement No 693024). FM is funded by a UKRI/EPSRC Stephen Hawking fellowship, grant reference EP/T017279/1. This work has been partially supported by STFC consolidated grant ST/P000681/1. We acknowledge ICTP-CINECA HPC collaboration for granting access to CINECA-Marconi Skylake partition. Computational resources were also provided by the ICTP on the local HPC facilities (Argo). This work was also performed using the Cambridge Service for Data Driven Discovery (CSD3), part of which is operated by the University of Cambridge Research Computing on behalf of the STFC DiRAC HPC Facility (www.dirac.ac.uk). The DiRAC component of CSD3 was funded by BEIS capital funding via STFC capital grants ST/P002307/1 and ST/R002452/1 and STFC operations grant ST/R00689X/1. DiRAC is part of the National e-Infrastructure. The authors also acknowledge the computer resources at SuperMUCNG and the technical support provided by the Leibniz Supercomputing Centre via PRACE Grant No. 2018194669.


\appendix

\section{Appendix: Code validation and convergence}
\label{sec-convergence}

To check convergence we perform the simulations with base resolutions for the half-box of $N=32^3$ (LR), $N=64^3$ (MR) and $N=128^3$ (HR), with the middle case corresponding to the results presented. In each case four 2:1 refinement levels are enforced to ensure that the initial oscillon is well resolved, and additional levels are added dynamically if collapse to a black hole occurs. 

\begin{figure}[]\centering
\includegraphics[scale=0.45]{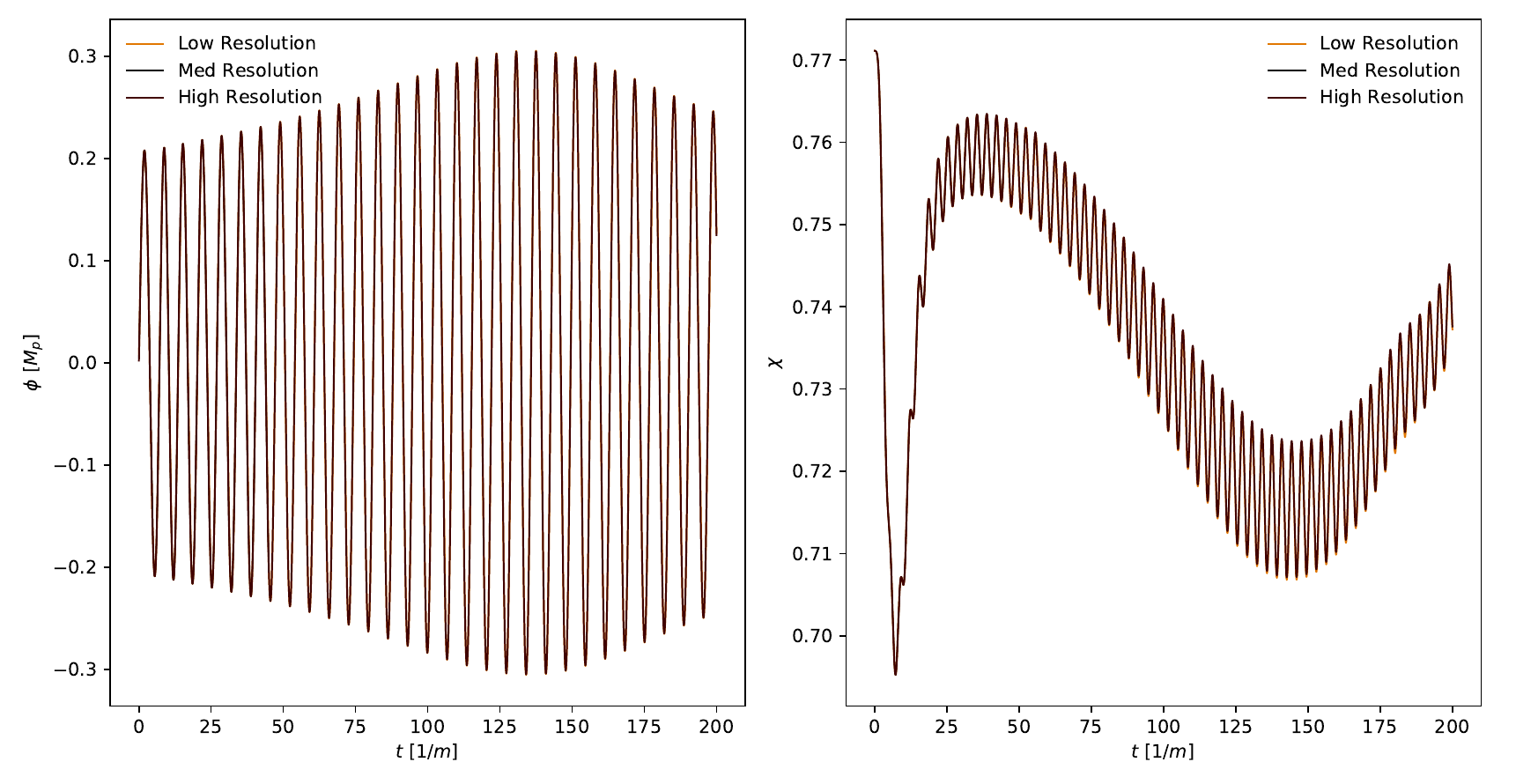}
\includegraphics[scale=0.45]{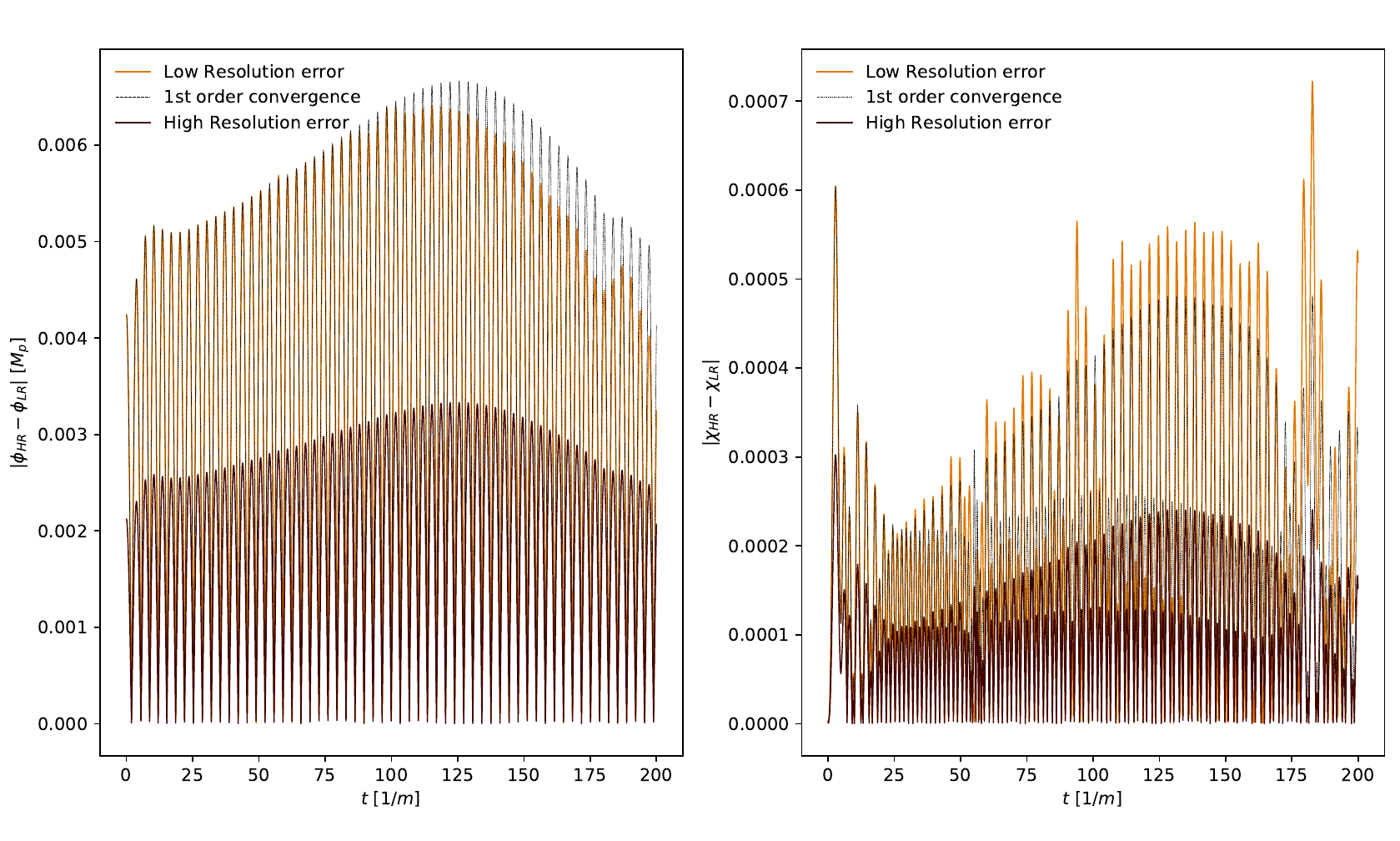}

\caption{Convergence test related to the case $q=1$, $\Lambda/M_p=1.$, $M = 2.63 \, M_p^2/m$.  The low resolution error is the absolute difference between the MR and LR values of the field at the centre of the grid, whereas the high resolution error is the absolute difference between the values at this point in the HR and MR cases. The first order convergence observed is consistent with the error introduced by the interpolation of our numerical solutions onto the initial grid. The ``first order convergence'' line is obtained by multiplying the high resolution error by a factor of 2, corresponding to the doubling in resolution, and should match (approximately) the magnitude of the low resolution error. In these plots we use the raw simulation time, which is in units of [1/m] and corresponds to the time measured by observers far from the oscillon.}
\label{fig-converge}
\end{figure}

We check that for the key quantities extracted at the centre of the grid, like the field $\phi$ and the conformal factor of the spatial metric $\chi$, numerical errors are not significant in the plots we present - see the top panel of Fig. \ref{fig-converge} for the actual values plotted at all three resolutions - we find that the differences are not visible to the naked eye. 

Our numerical integration scheme is between 3rd and 4th order accurate, but the results are dominated by the first order error from the interpolation of the numerical solutions onto the initial grid and our so convergence results are consistent with this, as shown in the bottom panel of Fig. \ref{fig-converge}. In future we plan to use a better interpolation scheme to improve the quality of the results.

We test convergence in all the cases either side of criticality - that is, in the highest mass case for which the oscillon is meta-stable, and in the lowest mass case for which a black hole is formed. In the cases where the solutions transitioned between stable and unstable at several masses, we performed additional checks and increased the (half) grid size from $L=64\, [1/m]$ to $L=128 \,[1/m]$ to check for boundary effects but found no evidence for errors of this kind affecting the outcomes. 

\begin{figure}[]\centering
\includegraphics[scale=0.45]{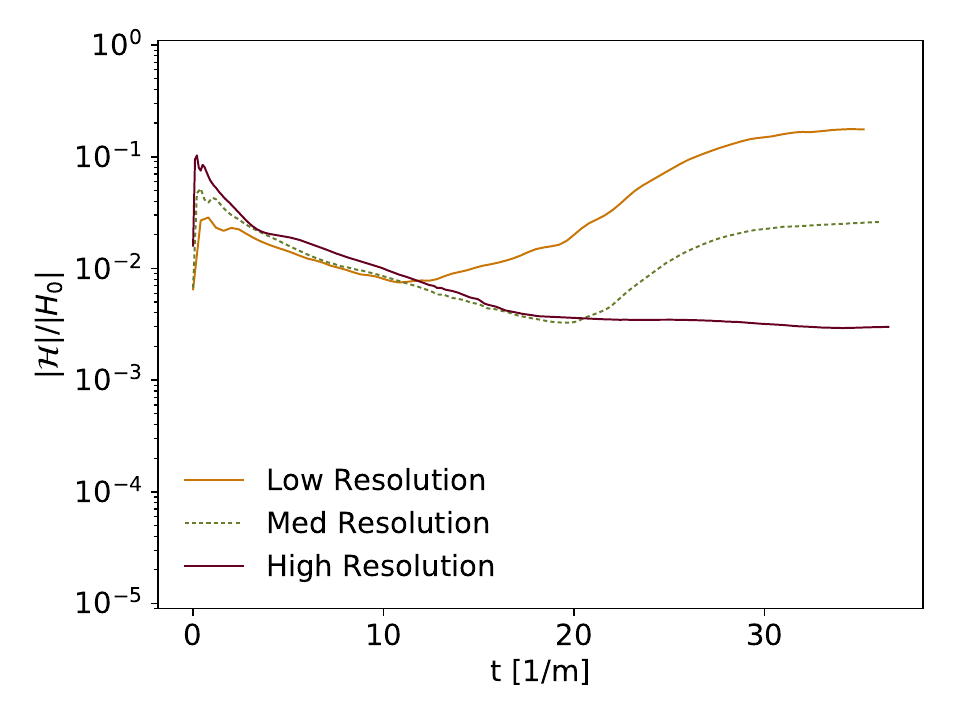}
\includegraphics[scale=0.45]{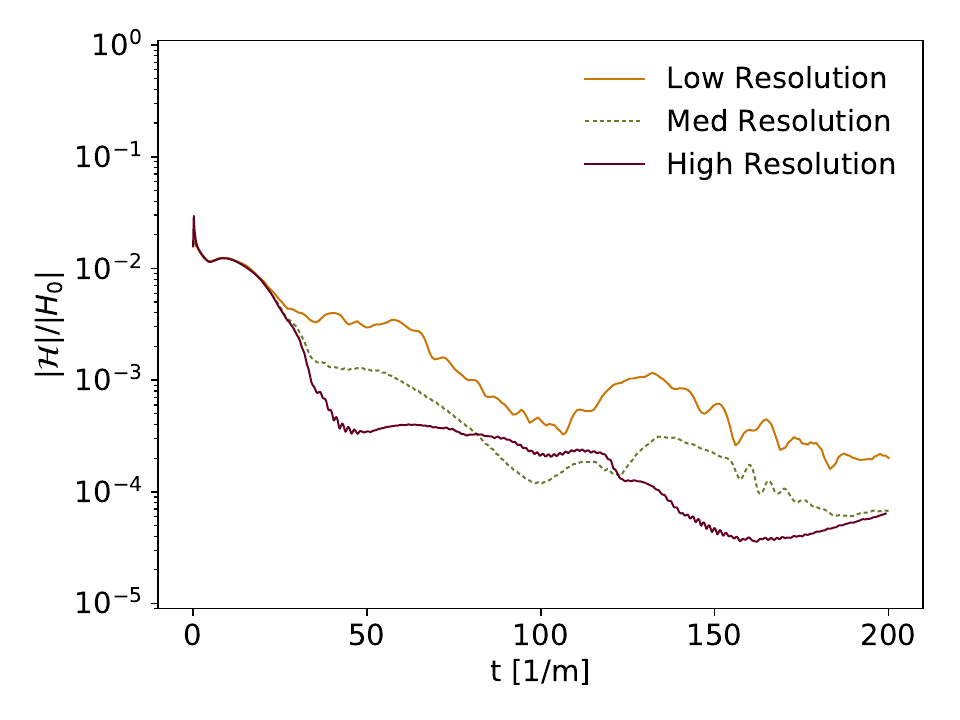}
\caption{The relative Hamiltonian constraint violation $|\mathcal{H}|/ H_0$ (as defined in the text) in a collapsing, BH forming case $q=0$, $\Lambda/M_p=1.$, $M = 3.01 \, M_p^2/m$ (left panel) and for a meta-stable case $q=0$, $\Lambda/M_p=1.$, $M = 2.07 \, M_p^2/m$ (right panel). The CCZ4 scheme is used to damp constraints, leading to an overall decrease over time in the meta-stable case (after an initial spike caused by the initial data - see the text for details). In the collapsing case there is a growth in the constraint violation as the BH forms (mainly due to regridding in the LR and MR cases) but this stabilises after the horizon forms (at around $t=30$ [1/m]), and converges away with increasing resolution.}
\label{fig-ham}
\end{figure}

We check that the Hamiltonian constraint violation in the simulations is reasonable in comparison to the typical energy density scale of the problem, and remains bounded with increasing resolution, as shown for two example cases in Fig. \ref{fig-ham}. The figure plotted is $|\mathcal{H}|/ |H_0|$ where (as in Eq. \eqref{eqn-Ham}) the Hamiltonian constraint is
\begin{equation}
	\mathcal{H} = {}^{(3)} R + K^2 + K_{ij} K^{ij} - 16 \pi \rho = 0 ~ .
\end{equation}
The norm $|\mathcal{H}|$ is taken as the volume integral of the absolute values of the Hamiltonian constraint over the portion of the grid within a coordinate distance of $r = 32$ [1/m] from the central point. In cases where a BH forms we excise the region within any horizon that forms when taking the integral, as it is known that large violations occur at the numerical BH ``puncture''. Such errors do not propagate outside of the horizon due to the causal structure of the spacetime. 

Since the value of the Hamiltonian constraint should be zero, a percentage error is not well defined. However, to give a guide as to how serious it is, one can compare it to the sum of the absolute values of the individual terms which occur in the Hamiltonian constraint (a kind of sum of total energy and curvature densities), thus we define
\begin{equation}
	H = |{}^{(3)} R | + K^2 + K_{ij} K^{ij} + 16 \pi \rho ~ ,
\end{equation}
with $H_0$ the initial value at $t=0$. The norm is taken over the same region as for $\mathcal{H}$ and therefore the quantity plotted $|\mathcal{H}|/ |H_0|$ can be roughly thought of as quantifying the error in the constraint in this region, relative to the initial total energy density.

The CCZ4 scheme is used to damp constraints, leading to an overall decrease over time in meta-stable cases. In the collapsing case there is a growth in the constraint violation as the BH forms (partly due to regridding errors in the MR and LR cases) but this stabilises after the horizon forms, and converges away with increasing resolution.

Whilst not clearly visible in Fig. \ref{fig-ham}, all resolutions start with approximately the same level of violation, at around 1-3\%. The value arises due to finite errors introduced by the conversion from areal polar to conformally flat initial data. This process involves integration of the areal polar data from a larger outer radius ($r=1000$ [1/m]) where the oscillon energy density is assumed to be negligible, inwards to the centre. It turns out that this process is highly sensitive to the small residual value of the field at this point, and so introduces the 1-3\% error which is then inherent in the data interpolated onto the grid. As such, whilst increasing the resolution reduces the gradient approximation error, this absolute error is not reduced (and thus the initial error does not always reduce). This initial error also creates a spike in the relative violation to around 3-5\% as it propagates out of the central region and crosses regridding boundaries, which is larger where the initial violation is larger. Given that the error remains relatively small throughout the simulations, that we get consistent results at all resolutions tested, and that we use the CCZ4 scheme to damp the error over time, we are confident that this does not affect our results. Nevertheless, in future investigations, we plan to directly interpolate the areal polar coordinate values, and adjust our gauge conditions to account for the non zero initial conformal flatness (which in their current form would introduce gauge artefacts).

\clearpage

\section{Field evolution for $q = -1$ cases}
\label{app:qminus1}
\begin{figure}[h!]\centering
\includegraphics[keepaspectratio, width=16cm, height=25cm]{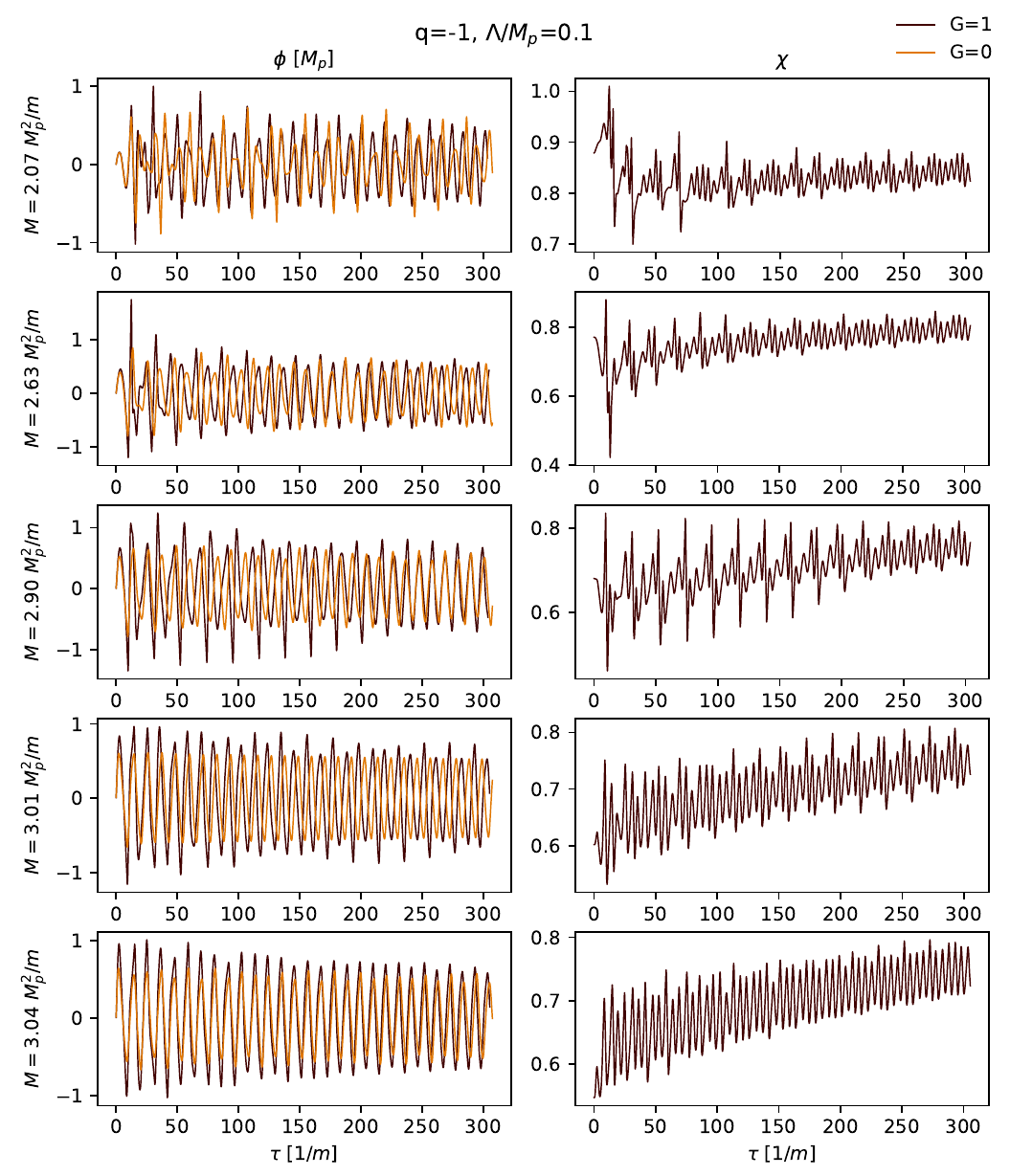}
\caption{The plots above show the full results for the case $q=-1$, $\Lambda=0.1 M_p$, with each row relating to a different total mass of the initial configuration as marked, and the left column showing the field value $\phi$ at the centre of the configuration, and the right column showing the value of the conformal factor of the metric $\chi$ at the same point.  The time shown is proper time for an observer at the centre in units of $1/m$.}
\end{figure}
\pagebreak
\begin{figure}[h!]\centering
\includegraphics[keepaspectratio, width=16cm, height=25cm]{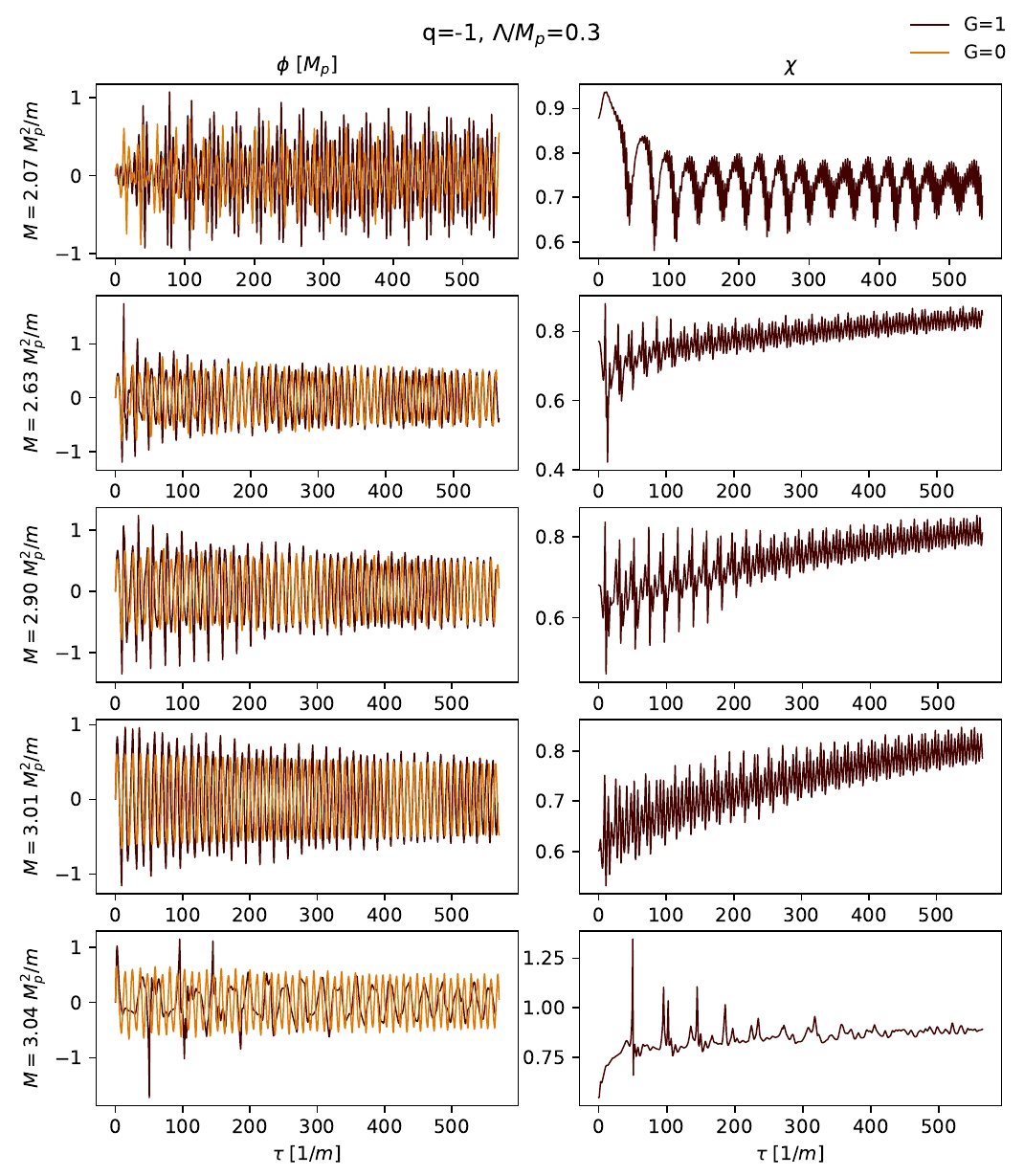}
\caption{The plots above show the full results for the case $q=-1$, $\Lambda=0.3 M_p$, with each row relating to a different total mass of the initial configuration as marked, and the left column showing the field value $\phi$ at the centre of the configuration, and the right column showing the value of the conformal factor of the metric $\chi$ at the same point. The time shown is proper time for an observer at the centre in units of $1/m$.}
\end{figure}
\pagebreak
\begin{figure}[h!]\centering
\includegraphics[keepaspectratio, width=16cm, height=25cm]{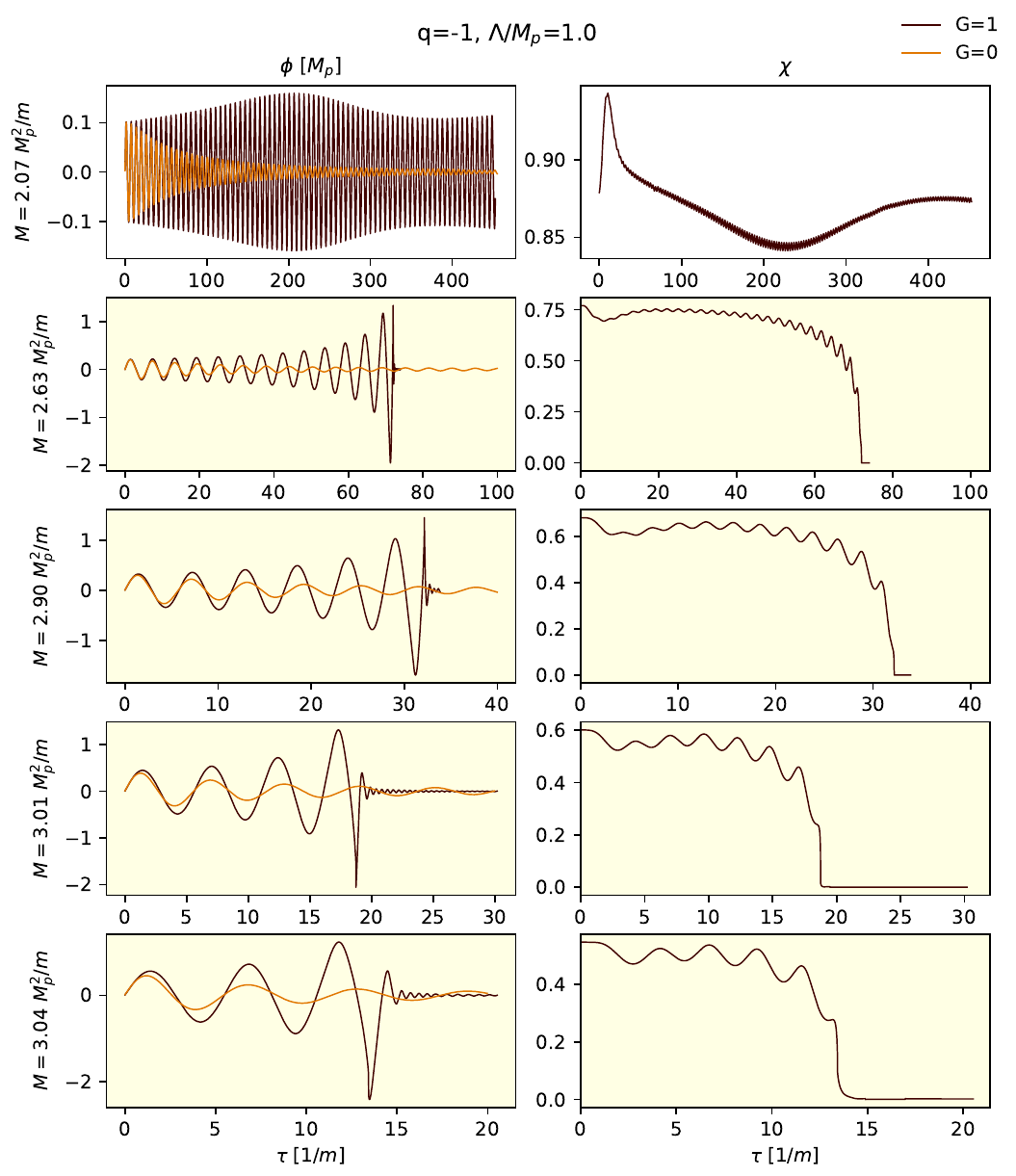}
\caption{The plots above show the full results for the case $q=-1$, $\Lambda=1.0 M_p$, with each row relating to a different total mass of the initial configuration as marked, and the left column showing the field value $\phi$ at the centre of the configuration, and the right column showing the value of the conformal factor of the metric $\chi$ at the same point. The time shown is proper time for an observer at the centre in units of $1/m$.}
\end{figure}

\clearpage
\section{Field evolution for $q = 0$ cases}

\label{app:q0}
\begin{figure}[h!]\centering
\includegraphics[keepaspectratio, width=16cm, height=25cm]{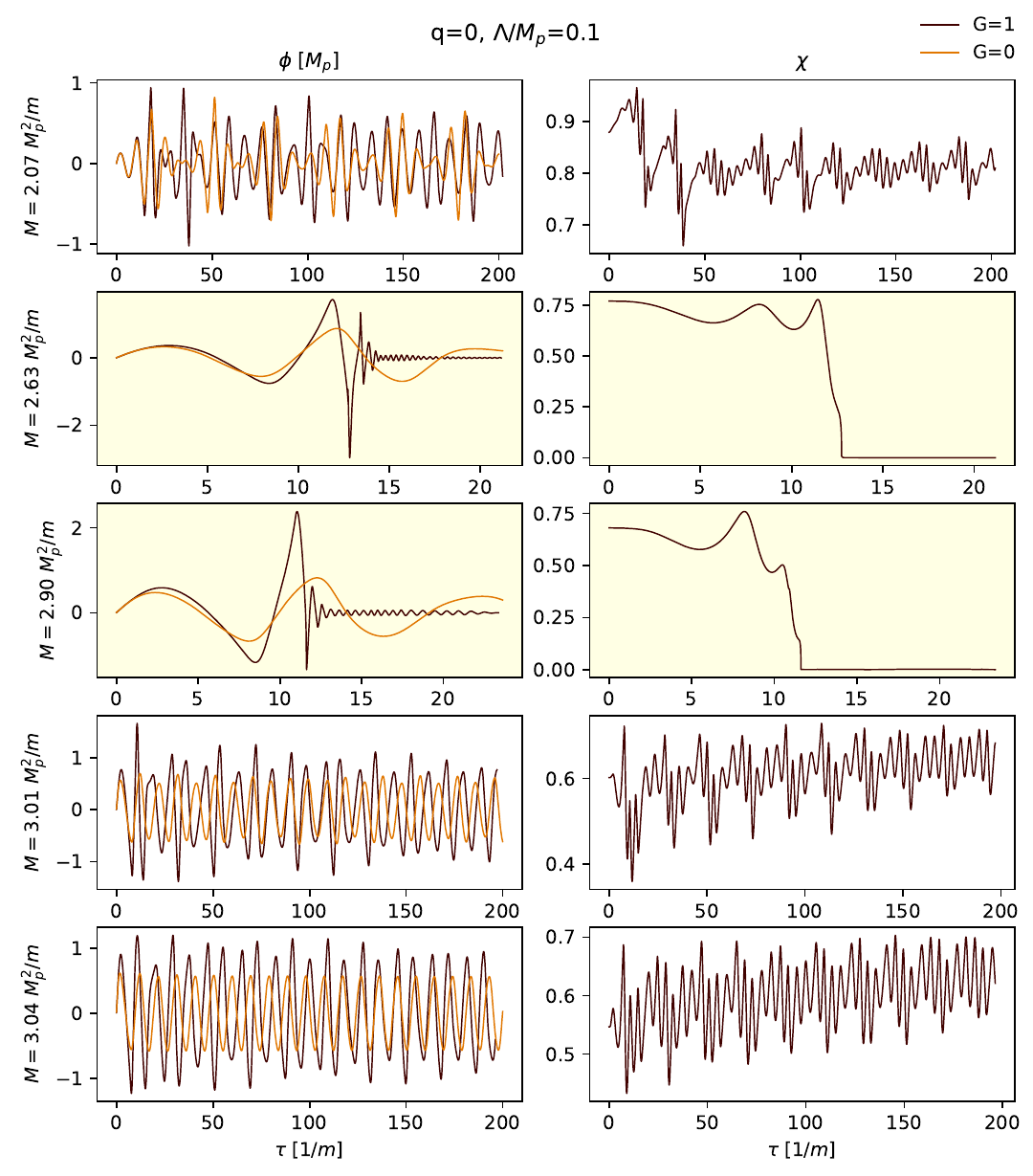}
\caption{The plots above show the full results for the case $q=0$, $\Lambda=0.1 M_p$, with each row relating to a different total mass of the initial configuration as marked, and the left column showing the field value $\phi$ at the centre of the configuration, and the right column showing the value of the conformal factor of the metric $\chi$ at the same point.  The time shown is proper time for an observer at the centre in units of $1/m$.}
\end{figure}
\pagebreak
\begin{figure}[h!]\centering
\includegraphics[keepaspectratio, width=16cm, height=25cm]{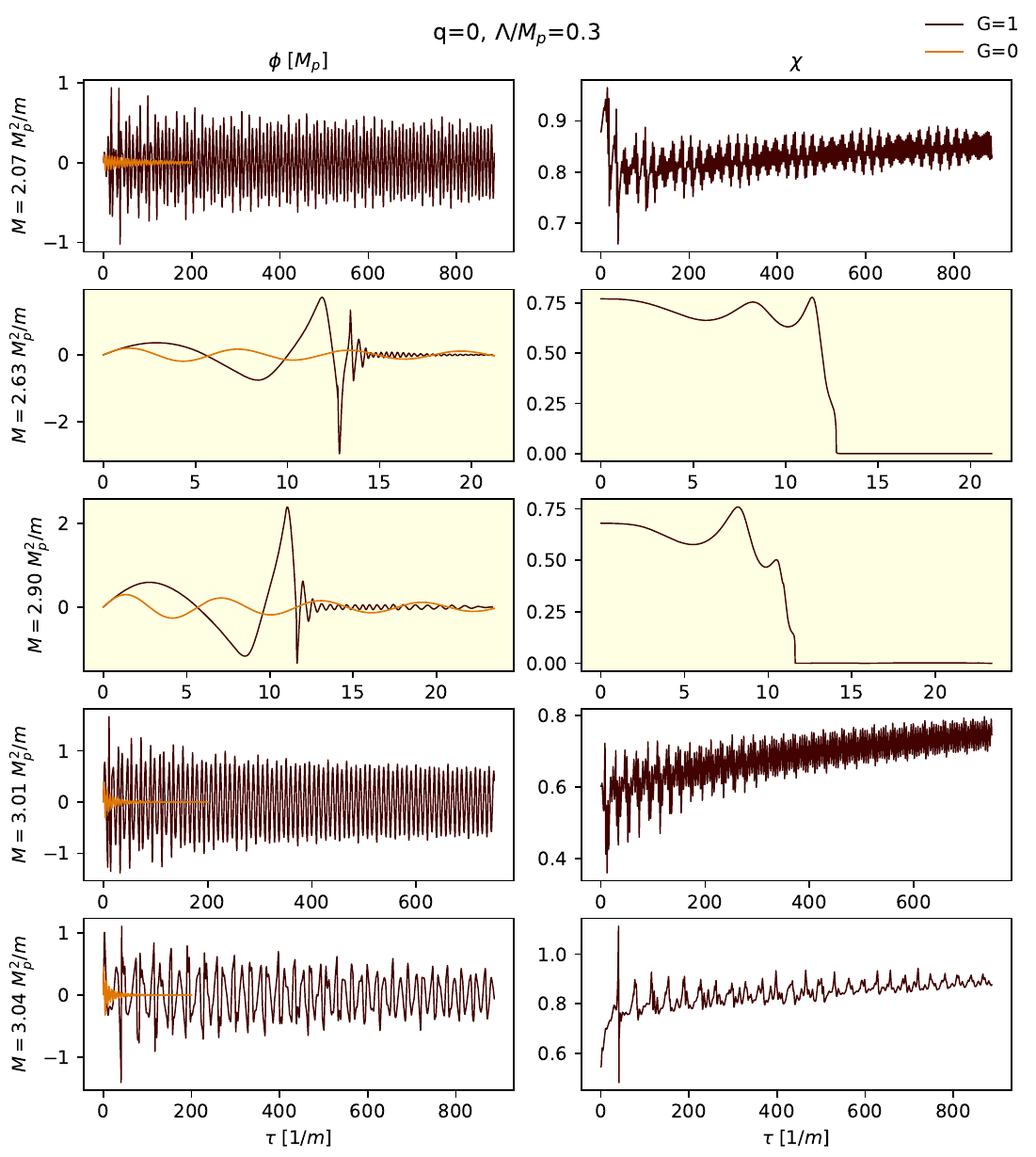}
\caption{The plots above show the full results for the case $q=0$, $\Lambda=0.3 M_p$, with each row relating to a different total mass of the initial configuration as marked, and the left column showing the field value $\phi$ at the centre of the configuration, and the right column showing the value of the conformal factor of the metric $\chi$ at the same point.  The time shown is proper time for an observer at the centre in units of $1/m$.}
\end{figure}
\pagebreak
\begin{figure}[h!]\centering
\includegraphics[keepaspectratio, width=16cm, height=25cm]{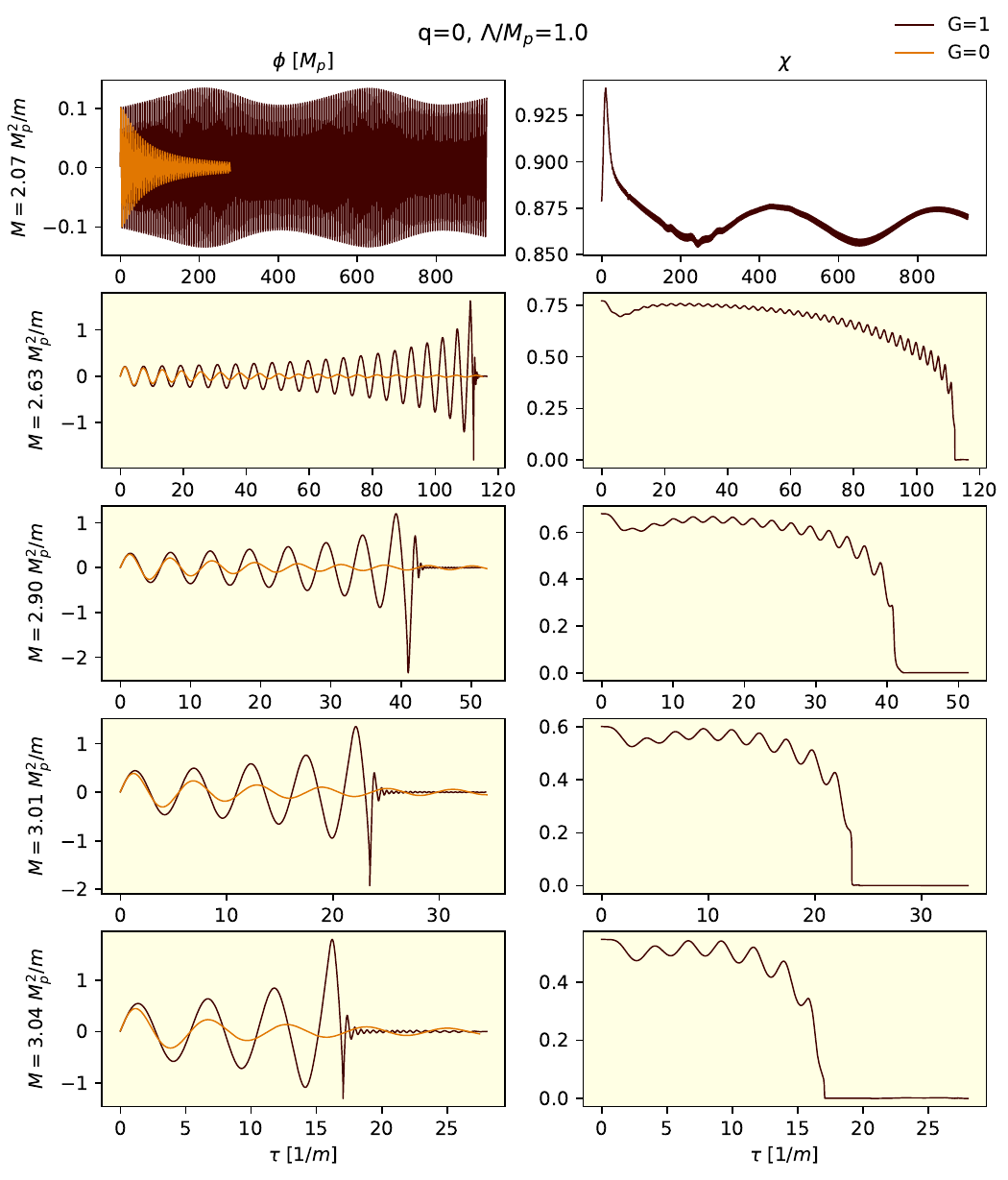}
\caption{The plots above show the full results for the case $q=0$, $\Lambda=1.0 M_p$, with each row relating to a different total mass of the initial configuration as marked, and the left column showing the field value $\phi$ at the centre of the configuration, and the right column showing the value of the conformal factor of the metric $\chi$ at the same point.  The time shown is proper time for an observer at the centre in units of $1/m$.}
\end{figure}
\pagebreak

\clearpage
\section{Field evolution for $q = 1$ cases}
\label{app:q1}

\begin{figure}[h!]\centering
\includegraphics[keepaspectratio, width=16cm, height=25cm]{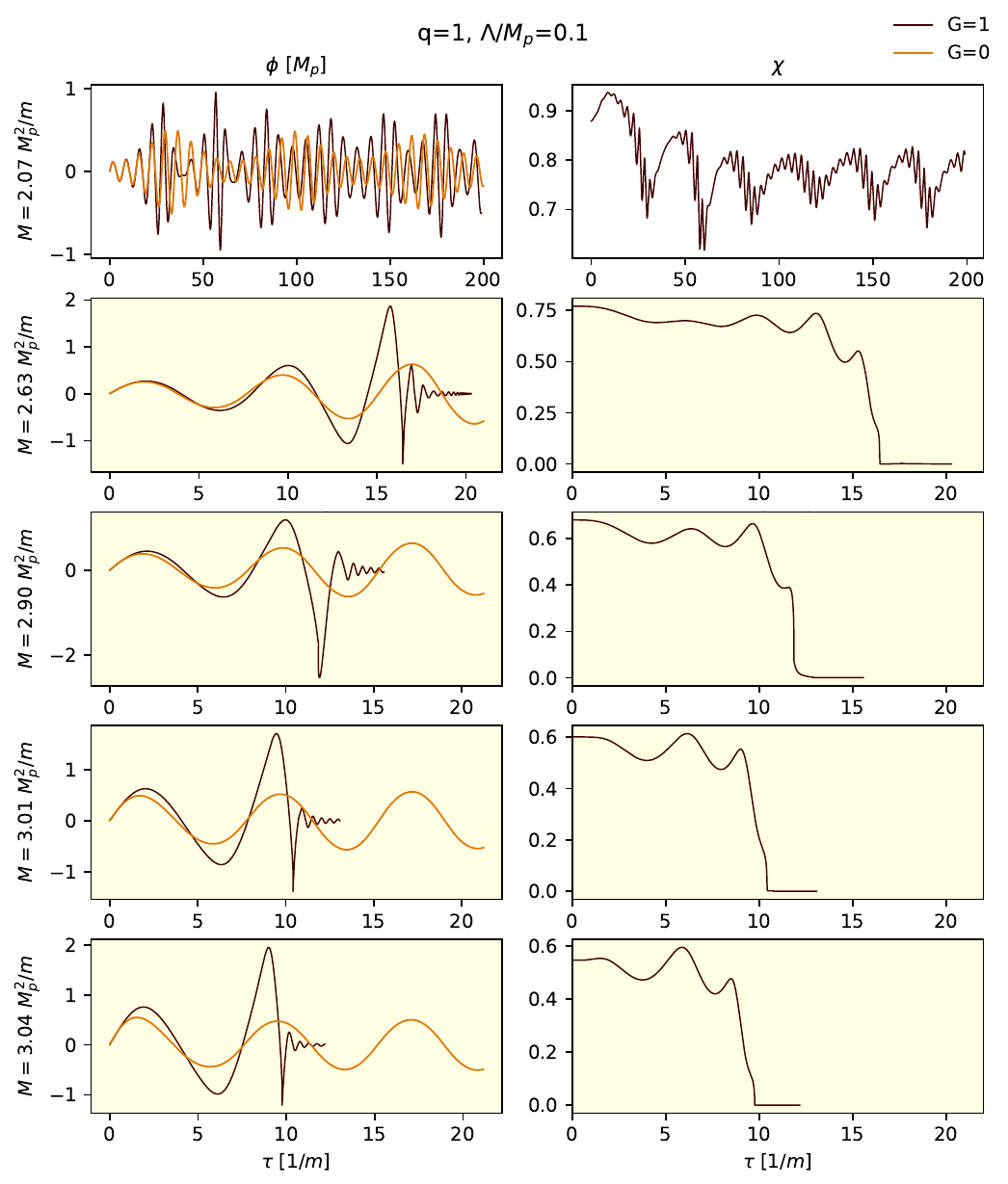}
\caption{The plots above show the full results for the case $q=1$, $\Lambda=0.1 M_p$, with each row relating to a different total mass of the initial configuration as marked, and the left column showing the field value $\phi$ at the centre of the configuration, and the right column showing the value of the conformal factor of the metric $\chi$ at the same point.  The time shown is proper time for an observer at the centre in units of $1/m$.}
\end{figure}
\pagebreak
\begin{figure}[h!]\centering
\includegraphics[keepaspectratio, width=16cm, height=25cm]{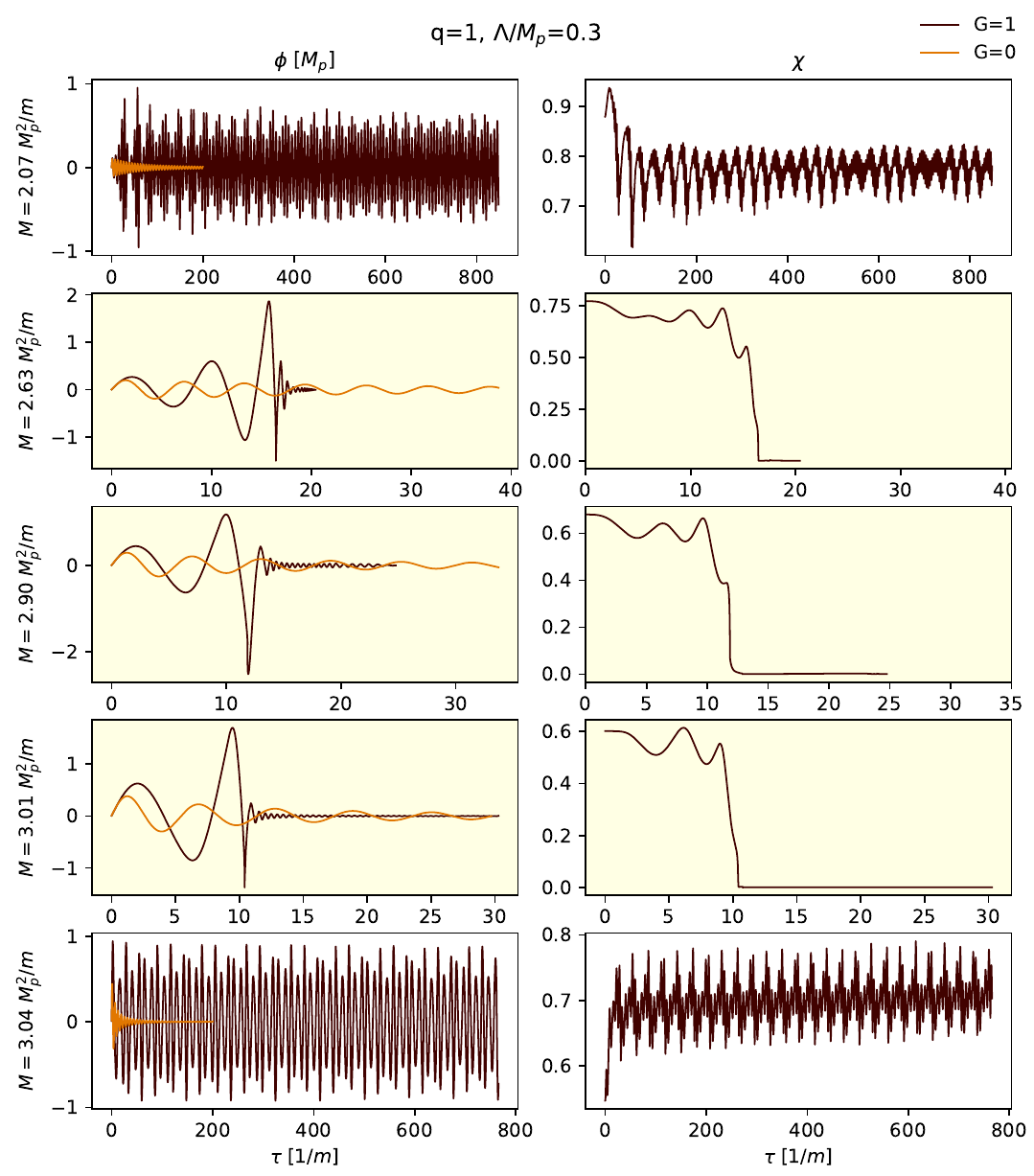}
\caption{The plots above show the full results for the case $q=1$, $\Lambda=0.3 M_p$, with each row relating to a different total mass of the initial configuration as marked, and the left column showing the field value $\phi$ at the centre of the configuration, and the right column showing the value of the conformal factor of the metric $\chi$ at the same point.  The time shown is proper time for an observer at the centre in units of $1/m$.}
\end{figure}
\pagebreak
\begin{figure}[h!]\centering
\includegraphics[keepaspectratio, width=16cm, height=25cm]{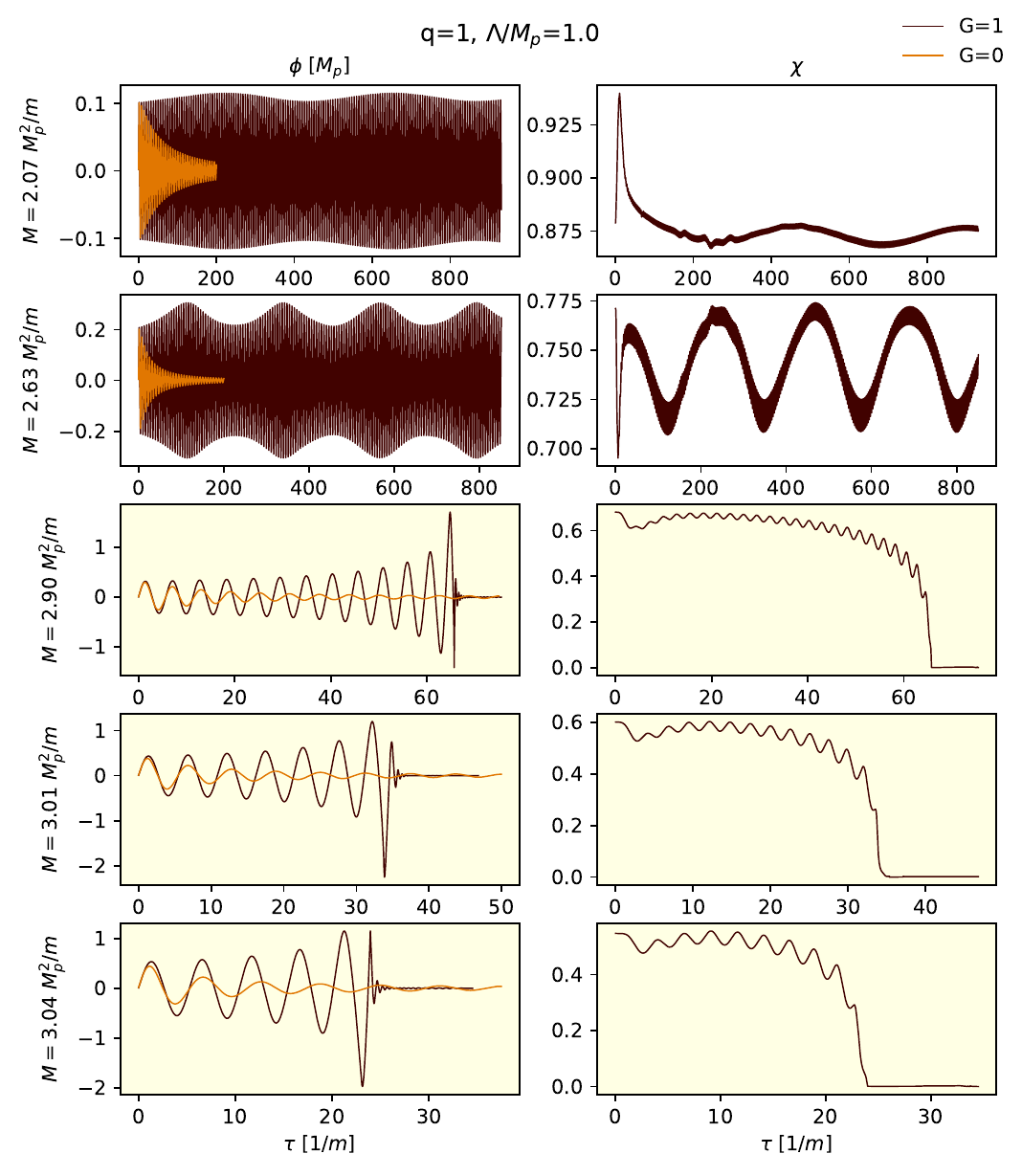}
\caption{The plots above show the full results for the case $q=1$, $\Lambda=1.0 M_p$, with each row relating to a different total mass of the initial configuration as marked, and the left column showing the field value $\phi$ at the centre of the configuration, and the right column showing the value of the conformal factor of the metric $\chi$ at the same point.  The time shown is proper time for an observer at the centre in units of $1/m$.}
\end{figure}

\medskip

%

\bibliographystyle{elsarticle-num}
\bibliography{axion}

\begin{thebibliography}{10}
\expandafter\ifx\csname url\endcsname\relax
  \def\url#1{\texttt{#1}}\fi
\expandafter\ifx\csname urlprefix\endcsname\relax\def\urlprefix{URL }\fi
\expandafter\ifx\csname href\endcsname\relax
  \def\href#1#2{#2} \def\path#1{#1}\fi

\bibitem{1006.3075}
M.~A. Amin, {Inflaton fragmentation: Emergence of pseudo-stable inflaton lumps
  (oscillons) after inflation} (2010).
\newblock \href {http://arxiv.org/abs/1006.3075} {\path{arXiv:1006.3075}}.

\bibitem{0712.3034}
E.~Farhi, N.~Graham, A.~H. Guth, N.~Iqbal, R.~R. Rosales, N.~Stamatopoulos,
  {Emergence of Oscillons in an Expanding Background}, Phys. Rev. D77 (2008)
  085019.
\newblock \href {http://arxiv.org/abs/0712.3034} {\path{arXiv:0712.3034}},
  \href {https://doi.org/10.1103/PhysRevD.77.085019}
  {\path{doi:10.1103/PhysRevD.77.085019}}.

\bibitem{1004.4658}
M.~Gleiser, N.~Graham, N.~Stamatopoulos, {Long-Lived Time-Dependent Remnants
  During Cosmological Symmetry Breaking: From Inflation to the Electroweak
  Scale}, Phys. Rev. D82 (2010) 043517.
\newblock \href {http://arxiv.org/abs/1004.4658} {\path{arXiv:1004.4658}},
  \href {https://doi.org/10.1103/PhysRevD.82.043517}
  {\path{doi:10.1103/PhysRevD.82.043517}}.

\bibitem{astro-ph/9311037}
E.~W. Kolb, I.~I. Tkachev, {Nonlinear axion dynamics and formation of
  cosmological pseudosolitons}, Phys. Rev. D49 (1994) 5040--5051.
\newblock \href {http://arxiv.org/abs/astro-ph/9311037}
  {\path{arXiv:astro-ph/9311037}}, \href
  {https://doi.org/10.1103/PhysRevD.49.5040}
  {\path{doi:10.1103/PhysRevD.49.5040}}.

\bibitem{1002.3380}
M.~A. Amin, D.~Shirokoff, {Flat-top oscillons in an expanding universe}, Phys.
  Rev. D81 (2010) 085045.
\newblock \href {http://arxiv.org/abs/1002.3380} {\path{arXiv:1002.3380}},
  \href {https://doi.org/10.1103/PhysRevD.81.085045}
  {\path{doi:10.1103/PhysRevD.81.085045}}.

\bibitem{1009.2505}
M.~A. Amin, R.~Easther, H.~Finkel, {Inflaton Fragmentation and Oscillon
  Formation in Three Dimensions}, JCAP 1012 (2010) 001.
\newblock \href {http://arxiv.org/abs/1009.2505} {\path{arXiv:1009.2505}},
  \href {https://doi.org/10.1088/1475-7516/2010/12/001}
  {\path{doi:10.1088/1475-7516/2010/12/001}}.

\bibitem{1103.1911}
M.~Gleiser, N.~Graham, N.~Stamatopoulos, {Generation of Coherent Structures
  After Cosmic Inflation}, Phys. Rev. D83 (2011) 096010.
\newblock \href {http://arxiv.org/abs/1103.1911} {\path{arXiv:1103.1911}},
  \href {https://doi.org/10.1103/PhysRevD.83.096010}
  {\path{doi:10.1103/PhysRevD.83.096010}}.

\bibitem{1106.3335}
M.~A. Amin, R.~Easther, H.~Finkel, R.~Flauger, M.~P. Hertzberg, {Oscillons
  After Inflation}, Phys. Rev. Lett. 108 (2012) 241302.
\newblock \href {http://arxiv.org/abs/1106.3335} {\path{arXiv:1106.3335}},
  \href {https://doi.org/10.1103/PhysRevLett.108.241302}
  {\path{doi:10.1103/PhysRevLett.108.241302}}.

\bibitem{1710.06851}
K.~D. Lozanov, M.~A. Amin, {Self-resonance after inflation: oscillons,
  transients and radiation domination}, Phys. Rev. D97~(2) (2018) 023533.
\newblock \href {http://arxiv.org/abs/1710.06851} {\path{arXiv:1710.06851}},
  \href {https://doi.org/10.1103/PhysRevD.97.023533}
  {\path{doi:10.1103/PhysRevD.97.023533}}.

\bibitem{Antusch:2017vga}
S.~Antusch, F.~Cefala, S.~Orani, {What can we learn from the stochastic
  gravitational wave background produced by oscillons?}, JCAP 03 (2018) 032.
\newblock \href {http://arxiv.org/abs/1712.03231} {\path{arXiv:1712.03231}},
  \href {https://doi.org/10.1088/1475-7516/2018/03/032}
  {\path{doi:10.1088/1475-7516/2018/03/032}}.

\bibitem{1711.10496}
J.-P. Hong, M.~Kawasaki, M.~Yamazaki, {Oscillons from Pure Natural Inflation},
  Phys. Rev. D98~(4) (2018) 043531.
\newblock \href {http://arxiv.org/abs/1711.10496} {\path{arXiv:1711.10496}},
  \href {https://doi.org/10.1103/PhysRevD.98.043531}
  {\path{doi:10.1103/PhysRevD.98.043531}}.

\bibitem{1902.07261}
M.~A. Amin, P.~Mocz, {Formation, gravitational clustering, and interactions of
  nonrelativistic solitons in an expanding universe}, Phys. Rev. D100~(6)
  (2019) 063507.
\newblock \href {http://arxiv.org/abs/1902.07261} {\path{arXiv:1902.07261}},
  \href {https://doi.org/10.1103/PhysRevD.100.063507}
  {\path{doi:10.1103/PhysRevD.100.063507}}.

\bibitem{1902.06736}
K.~D. Lozanov, M.~A. Amin, {Gravitational perturbations from oscillons and
  transients after inflation}, Phys. Rev. D99~(12) (2019) 123504.
\newblock \href {http://arxiv.org/abs/1902.06736} {\path{arXiv:1902.06736}},
  \href {https://doi.org/10.1103/PhysRevD.99.123504}
  {\path{doi:10.1103/PhysRevD.99.123504}}.

\bibitem{Sang:2019ndv}
Y.~Sang, Q.-G. Huang, {Stochastic Gravitational-Wave Background from
  Axion-Monodromy Oscillons in String Theory During Preheating}, Phys. Rev. D
  100~(6) (2019) 063516.
\newblock \href {http://arxiv.org/abs/1905.00371} {\path{arXiv:1905.00371}},
  \href {https://doi.org/10.1103/PhysRevD.100.063516}
  {\path{doi:10.1103/PhysRevD.100.063516}}.

\bibitem{1304.6094}
S.-Y. Zhou, E.~J. Copeland, R.~Easther, H.~Finkel, Z.-G. Mou, P.~M. Saffin,
  {Gravitational Waves from Oscillon Preheating}, JHEP 10 (2013) 026.
\newblock \href {http://arxiv.org/abs/1304.6094} {\path{arXiv:1304.6094}},
  \href {https://doi.org/10.1007/JHEP10(2013)026}
  {\path{doi:10.1007/JHEP10(2013)026}}.

\bibitem{1607.01314}
S.~Antusch, F.~Cefala, S.~Orani, {Gravitational waves from oscillons after
  inflation}, Phys. Rev. Lett. 118~(1) (2017) 011303, [Erratum: Phys. Rev.
  Lett.120,no.21,219901(2018)].
\newblock \href {http://arxiv.org/abs/1607.01314} {\path{arXiv:1607.01314}},
  \href {https://doi.org/10.1103/PhysRevLett.120.219901,
  10.1103/PhysRevLett.118.011303} {\path{doi:10.1103/PhysRevLett.120.219901,
  10.1103/PhysRevLett.118.011303}}.

\bibitem{1707.09841}
J.~Liu, Z.-K. Guo, R.-G. Cai, G.~Shiu, {Gravitational Waves from Oscillons with
  Cuspy Potentials}, Phys. Rev. Lett. 120~(3) (2018) 031301.
\newblock \href {http://arxiv.org/abs/1707.09841} {\path{arXiv:1707.09841}},
  \href {https://doi.org/10.1103/PhysRevLett.120.031301}
  {\path{doi:10.1103/PhysRevLett.120.031301}}.

\bibitem{1708.08922}
S.~Antusch, F.~Cefala, S.~Krippendorf, F.~Muia, S.~Orani, F.~Quevedo,
  {Oscillons from String Moduli}, JHEP 01 (2018) 083.
\newblock \href {http://arxiv.org/abs/1708.08922} {\path{arXiv:1708.08922}},
  \href {https://doi.org/10.1007/JHEP01(2018)083}
  {\path{doi:10.1007/JHEP01(2018)083}}.

\bibitem{1803.08047}
M.~A. Amin, J.~Braden, E.~J. Copeland, J.~T. Giblin, C.~Solorio, Z.~J. Weiner,
  S.-Y. Zhou, {Gravitational waves from asymmetric oscillon dynamics?}, Phys.
  Rev. D98 (2018) 024040.
\newblock \href {http://arxiv.org/abs/1803.08047} {\path{arXiv:1803.08047}},
  \href {https://doi.org/10.1103/PhysRevD.98.024040}
  {\path{doi:10.1103/PhysRevD.98.024040}}.

\bibitem{hep-ph/9303313}
E.~W. Kolb, I.~I. Tkachev, {Axion miniclusters and Bose stars}, Phys. Rev.
  Lett. 71 (1993) 3051--3054.
\newblock \href {http://arxiv.org/abs/hep-ph/9303313}
  {\path{arXiv:hep-ph/9303313}}, \href
  {https://doi.org/10.1103/PhysRevLett.71.3051}
  {\path{doi:10.1103/PhysRevLett.71.3051}}.

\bibitem{1906.06352}
J.~Ollé, O.~Pujolàs, F.~Rompineve, {Oscillons and Dark Matter}, JCAP 2002
  (2020) 006.
\newblock \href {http://arxiv.org/abs/1906.06352} {\path{arXiv:1906.06352}},
  \href {https://doi.org/10.1088/1475-7516/2020/02/006}
  {\path{doi:10.1088/1475-7516/2020/02/006}}.

\bibitem{1909.11665}
A.~Arvanitaki, S.~Dimopoulos, M.~Galanis, L.~Lehner, J.~O. Thompson,
  K.~Van~Tilburg, {Large-misalignment mechanism for the formation of compact
  axion structures: Signatures from the QCD axion to fuzzy dark matter}, Phys.
  Rev. D101~(8) (2020) 083014.
\newblock \href {http://arxiv.org/abs/1909.11665} {\path{arXiv:1909.11665}},
  \href {https://doi.org/10.1103/PhysRevD.101.083014}
  {\path{doi:10.1103/PhysRevD.101.083014}}.

\bibitem{1909.10805}
M.~Kawasaki, W.~Nakano, E.~Sonomoto, {Oscillon of Ultra-Light Axion-like
  Particle}, JCAP 2001 (2020) 047.
\newblock \href {http://arxiv.org/abs/1909.10805} {\path{arXiv:1909.10805}},
  \href {https://doi.org/10.1088/1475-7516/2020/01/047}
  {\path{doi:10.1088/1475-7516/2020/01/047}}.

\bibitem{1912.07064}
J.~C. Niemeyer, {Small-scale structure of fuzzy and axion-like dark matter},
  Prog. Part. Nucl. Phys. (2019) 103787\href {http://arxiv.org/abs/1912.07064}
  {\path{arXiv:1912.07064}}, \href {https://doi.org/10.1016/j.ppnp.2020.103787}
  {\path{doi:10.1016/j.ppnp.2020.103787}}.

\bibitem{Boskovic:2018rub}
M.~Bo\v{s}kovi\'c, F.~Duque, M.~C. Ferreira, F.~S. Miguel, V.~Cardoso, {Motion
  in time-periodic backgrounds with applications to ultralight dark matter
  haloes at galactic centers}, Phys. Rev. D 98 (2018) 024037.
\newblock \href {http://arxiv.org/abs/1806.07331} {\path{arXiv:1806.07331}},
  \href {https://doi.org/10.1103/PhysRevD.98.024037}
  {\path{doi:10.1103/PhysRevD.98.024037}}.

\bibitem{Ferreira:2017pth}
M.~C. Ferreira, C.~F.~B. Macedo, V.~Cardoso, {Orbital fingerprints of
  ultralight scalar fields around black holes}, Phys. Rev. D 96~(8) (2017)
  083017.
\newblock \href {http://arxiv.org/abs/1710.00830} {\path{arXiv:1710.00830}},
  \href {https://doi.org/10.1103/PhysRevD.96.083017}
  {\path{doi:10.1103/PhysRevD.96.083017}}.

\bibitem{Annulli:2020lyc}
L.~Annulli, V.~Cardoso, R.~Vicente, {Response of ultralight dark matter to
  supermassive black holes and binaries}, Phys. Rev. D 102~(6) (2020) 063022.
\newblock \href {http://arxiv.org/abs/2009.00012} {\path{arXiv:2009.00012}},
  \href {https://doi.org/10.1103/PhysRevD.102.063022}
  {\path{doi:10.1103/PhysRevD.102.063022}}.

\bibitem{Annulli:2020ilw}
L.~Annulli, V.~Cardoso, R.~Vicente, {Stirred and shaken: dynamical behavior of
  boson stars and dark matter cores} (7 2020).
\newblock \href {http://arxiv.org/abs/2007.03700} {\path{arXiv:2007.03700}}.

\bibitem{Khmelnitsky:2013lxt}
A.~Khmelnitsky, V.~Rubakov, {Pulsar timing signal from ultralight scalar dark
  matter}, JCAP 02 (2014) 019.
\newblock \href {http://arxiv.org/abs/1309.5888} {\path{arXiv:1309.5888}},
  \href {https://doi.org/10.1088/1475-7516/2014/02/019}
  {\path{doi:10.1088/1475-7516/2014/02/019}}.

\bibitem{1408.1811}
K.~D. Lozanov, M.~A. Amin, {End of inflation, oscillons, and matter-antimatter
  asymmetry}, Phys. Rev. D90~(8) (2014) 083528.
\newblock \href {http://arxiv.org/abs/1408.1811} {\path{arXiv:1408.1811}},
  \href {https://doi.org/10.1103/PhysRevD.90.083528}
  {\path{doi:10.1103/PhysRevD.90.083528}}.

\bibitem{Bogolyubsky:1976yu}
I.~L. Bogolyubsky, V.~G. Makhankov, {Lifetime of Pulsating Solitons in Some
  Classical Models}, Pisma Zh. Eksp. Teor. Fiz. 24 (1976) 15--18.

\bibitem{hep-ph/9308279}
M.~Gleiser, {Pseudostable bubbles}, Phys. Rev. D49 (1994) 2978--2981.
\newblock \href {http://arxiv.org/abs/hep-ph/9308279}
  {\path{arXiv:hep-ph/9308279}}, \href
  {https://doi.org/10.1103/PhysRevD.49.2978}
  {\path{doi:10.1103/PhysRevD.49.2978}}.

\bibitem{hep-ph/9503217}
E.~J. Copeland, M.~Gleiser, H.~R. Muller, {Oscillons: Resonant configurations
  during bubble collapse}, Phys. Rev. D52 (1995) 1920--1933.
\newblock \href {http://arxiv.org/abs/hep-ph/9503217}
  {\path{arXiv:hep-ph/9503217}}, \href
  {https://doi.org/10.1103/PhysRevD.52.1920}
  {\path{doi:10.1103/PhysRevD.52.1920}}.

\bibitem{hep-ph/0209358}
S.~Kasuya, M.~Kawasaki, F.~Takahashi, {I-balls}, Phys. Lett. B559 (2003)
  99--106.
\newblock \href {http://arxiv.org/abs/hep-ph/0209358}
  {\path{arXiv:hep-ph/0209358}}, \href
  {https://doi.org/10.1016/S0370-2693(03)00344-7}
  {\path{doi:10.1016/S0370-2693(03)00344-7}}.

\bibitem{1806.04690}
S.~Krippendorf, F.~Muia, F.~Quevedo, {Moduli Stars}, JHEP 08 (2018) 070.
\newblock \href {http://arxiv.org/abs/1806.04690} {\path{arXiv:1806.04690}},
  \href {https://doi.org/10.1007/JHEP08(2018)070}
  {\path{doi:10.1007/JHEP08(2018)070}}.

\bibitem{Segur:1987mg}
H.~Segur, M.~D. Kruskal, {Nonexistence of Small Amplitude Breather Solutions in
  $\phi^4$ Theory}, Phys. Rev. Lett. 58 (1987) 747--750.
\newblock \href {https://doi.org/10.1103/PhysRevLett.58.747}
  {\path{doi:10.1103/PhysRevLett.58.747}}.

\bibitem{gr-qc/0301105}
M.~Alcubierre, R.~Becerril, S.~F. Guzman, T.~Matos, D.~Nunez, L.~A.
  Urena-Lopez, {Numerical studies of $\Phi^2$ oscillatons}, Class. Quant. Grav.
  20 (2003) 2883--2904.
\newblock \href {http://arxiv.org/abs/gr-qc/0301105}
  {\path{arXiv:gr-qc/0301105}}, \href
  {https://doi.org/10.1088/0264-9381/20/13/332}
  {\path{doi:10.1088/0264-9381/20/13/332}}.

\bibitem{gr-qc/0104093}
L.~A. Urena-Lopez, {Oscillatons revisited}, Class. Quant. Grav. 19 (2002)
  2617--2632.
\newblock \href {http://arxiv.org/abs/gr-qc/0104093}
  {\path{arXiv:gr-qc/0104093}}, \href
  {https://doi.org/10.1088/0264-9381/19/10/307}
  {\path{doi:10.1088/0264-9381/19/10/307}}.

\bibitem{UrenaLopez:2002gx}
L.~A. Urena-Lopez, T.~Matos, R.~Becerril, {Inside oscillatons}, Class. Quant.
  Grav. 19 (2002) 6259--6277.
\newblock \href {https://doi.org/10.1088/0264-9381/19/23/320}
  {\path{doi:10.1088/0264-9381/19/23/320}}.

\bibitem{UrenaLopez:2012zz}
L.~A. Urena-Lopez, S.~Valdez-Alvarado, R.~Becerril, {Evolution and stability
  $\phi^4$ oscillatons}, Class. Quant. Grav. 29 (2012) 065021.
\newblock \href {https://doi.org/10.1088/0264-9381/29/6/065021}
  {\path{doi:10.1088/0264-9381/29/6/065021}}.

\bibitem{Coleman:1985ki}
S.~R. Coleman, {Q Balls}, Nucl. Phys. B262 (1985) 263, [Erratum: Nucl.
  Phys.B269,744(1986)].
\newblock \href {https://doi.org/10.1016/0550-3213(85)90286-X,
  10.1016/0550-3213(86)90520-1} {\path{doi:10.1016/0550-3213(85)90286-X,
  10.1016/0550-3213(86)90520-1}}.

\bibitem{Jetzer:1991jr}
P.~Jetzer, {Boson stars}, Phys. Rept. 220 (1992) 163--227.
\newblock \href {https://doi.org/10.1016/0370-1573(92)90123-H}
  {\path{doi:10.1016/0370-1573(92)90123-H}}.

\bibitem{hep-th/0610267}
N.~Graham, {An Electroweak oscillon}, Phys. Rev. Lett. 98 (2007) 101801,
  [Erratum: Phys. Rev. Lett.98,189904(2007)].
\newblock \href {http://arxiv.org/abs/hep-th/0610267}
  {\path{arXiv:hep-th/0610267}}, \href
  {https://doi.org/10.1103/PhysRevLett.98.101801,
  10.1103/PhysRevLett.98.189904} {\path{doi:10.1103/PhysRevLett.98.101801,
  10.1103/PhysRevLett.98.189904}}.

\bibitem{0808.0514}
M.~Gleiser, J.~Thorarinson, {A Class of Nonperturbative Configurations in
  Abelian-Higgs Models: Complexity from Dynamical Symmetry Breaking}, Phys.
  Rev. D79 (2009) 025016.
\newblock \href {http://arxiv.org/abs/0808.0514} {\path{arXiv:0808.0514}},
  \href {https://doi.org/10.1103/PhysRevD.79.025016}
  {\path{doi:10.1103/PhysRevD.79.025016}}.

\bibitem{1210.7568}
E.~I. Sfakianakis, {Analysis of Oscillons in the SU(2) Gauged Higgs Model}
  (2012).
\newblock \href {http://arxiv.org/abs/1210.7568} {\path{arXiv:1210.7568}}.

\bibitem{Antusch:2015ziz}
S.~Antusch, S.~Orani, {Impact of other scalar fields on oscillons after hilltop
  inflation}, JCAP 03 (2016) 026.
\newblock \href {http://arxiv.org/abs/1511.02336} {\path{arXiv:1511.02336}},
  \href {https://doi.org/10.1088/1475-7516/2016/03/026}
  {\path{doi:10.1088/1475-7516/2016/03/026}}.

\bibitem{Amin:2013ika}
M.~A. Amin, {K-oscillons: Oscillons with noncanonical kinetic terms}, Phys.
  Rev. D87~(12) (2013) 123505.
\newblock \href {http://arxiv.org/abs/1303.1102} {\path{arXiv:1303.1102}},
  \href {https://doi.org/10.1103/PhysRevD.87.123505}
  {\path{doi:10.1103/PhysRevD.87.123505}}.

\bibitem{1809.07724}
J.~Sakstein, M.~Trodden, {Oscillons in Higher-Derivative Effective Field
  Theories}, Phys. Rev. D98~(12) (2018) 123512.
\newblock \href {http://arxiv.org/abs/1809.07724} {\path{arXiv:1809.07724}},
  \href {https://doi.org/10.1103/PhysRevD.98.123512}
  {\path{doi:10.1103/PhysRevD.98.123512}}.

\bibitem{Piette:1997hf}
B.~Piette, W.~J. Zakrzewski, {Metastable stationary solutions of the radial
  d-dimensional sine-Gordon model}, Nonlinearity 11 (1998) 1103--1110.
\newblock \href {https://doi.org/10.1088/0951-7715/11/4/020}
  {\path{doi:10.1088/0951-7715/11/4/020}}.

\bibitem{0812.1919}
G.~Fodor, P.~Forgacs, Z.~Horvath, M.~Mezei, {Computation of the radiation
  amplitude of oscillons}, Phys. Rev. D79 (2009) 065002.
\newblock \href {http://arxiv.org/abs/0812.1919} {\path{arXiv:0812.1919}},
  \href {https://doi.org/10.1103/PhysRevD.79.065002}
  {\path{doi:10.1103/PhysRevD.79.065002}}.

\bibitem{0903.0953}
G.~Fodor, P.~Forgacs, Z.~Horvath, M.~Mezei, {Radiation of scalar oscillons in 2
  and 3 dimensions}, Phys. Lett. B674 (2009) 319--324.
\newblock \href {http://arxiv.org/abs/0903.0953} {\path{arXiv:0903.0953}},
  \href {https://doi.org/10.1016/j.physletb.2009.03.054}
  {\path{doi:10.1016/j.physletb.2009.03.054}}.

\bibitem{0910.5922}
M.~Gleiser, D.~Sicilia, {A General Theory of Oscillon Dynamics}, Phys. Rev. D80
  (2009) 125037.
\newblock \href {http://arxiv.org/abs/0910.5922} {\path{arXiv:0910.5922}},
  \href {https://doi.org/10.1103/PhysRevD.80.125037}
  {\path{doi:10.1103/PhysRevD.80.125037}}.

\bibitem{1003.3459}
M.~P. Hertzberg, {Quantum Radiation of Oscillons}, Phys. Rev. D82 (2010)
  045022.
\newblock \href {http://arxiv.org/abs/1003.3459} {\path{arXiv:1003.3459}},
  \href {https://doi.org/10.1103/PhysRevD.82.045022}
  {\path{doi:10.1103/PhysRevD.82.045022}}.

\bibitem{1201.1934}
P.~Salmi, M.~Hindmarsh, {Radiation and Relaxation of Oscillons}, Phys. Rev. D85
  (2012) 085033.
\newblock \href {http://arxiv.org/abs/1201.1934} {\path{arXiv:1201.1934}},
  \href {https://doi.org/10.1103/PhysRevD.85.085033}
  {\path{doi:10.1103/PhysRevD.85.085033}}.

\bibitem{1210.2227}
E.~A. Andersen, A.~Tranberg, {Four results on $phi^4$ oscillons in D+1
  dimensions}, JHEP 12 (2012) 016.
\newblock \href {http://arxiv.org/abs/1210.2227} {\path{arXiv:1210.2227}},
  \href {https://doi.org/10.1007/JHEP12(2012)016}
  {\path{doi:10.1007/JHEP12(2012)016}}.

\bibitem{1612.07750}
K.~Mukaida, M.~Takimoto, M.~Yamada, {On Longevity of I-ball/Oscillon}, JHEP 03
  (2017) 122.
\newblock \href {http://arxiv.org/abs/1612.07750} {\path{arXiv:1612.07750}},
  \href {https://doi.org/10.1007/JHEP03(2017)122}
  {\path{doi:10.1007/JHEP03(2017)122}}.

\bibitem{1901.06130}
M.~Ibe, M.~Kawasaki, W.~Nakano, E.~Sonomoto, {Decay of I-ball/Oscillon in
  Classical Field Theory}, JHEP 04 (2019) 030.
\newblock \href {http://arxiv.org/abs/1901.06130} {\path{arXiv:1901.06130}},
  \href {https://doi.org/10.1007/JHEP04(2019)030}
  {\path{doi:10.1007/JHEP04(2019)030}}.

\bibitem{1906.04070}
M.~Gleiser, M.~Krackow, {Resonant configurations in scalar field theories: Can
  some oscillons live forever?}, Phys. Rev. D100~(11) (2019) 116005.
\newblock \href {http://arxiv.org/abs/1906.04070} {\path{arXiv:1906.04070}},
  \href {https://doi.org/10.1103/PhysRevD.100.116005}
  {\path{doi:10.1103/PhysRevD.100.116005}}.

\bibitem{Antusch:2019qrr}
S.~Antusch, F.~Cefal\`a, F.~Torrent\'\i{}, {Properties of Oscillons in Hilltop
  Potentials: energies, shapes, and lifetimes}, JCAP 10 (2019) 002.
\newblock \href {http://arxiv.org/abs/1907.00611} {\path{arXiv:1907.00611}},
  \href {https://doi.org/10.1088/1475-7516/2019/10/002}
  {\path{doi:10.1088/1475-7516/2019/10/002}}.

\bibitem{1908.11103}
M.~Ibe, M.~Kawasaki, W.~Nakano, E.~Sonomoto, {Fragileness of Exact
  I-ball/Oscillon}, Phys. Rev. D100~(12) (2020) 125021, [Phys.
  Rev.D100,125021(2019)].
\newblock \href {http://arxiv.org/abs/1908.11103} {\path{arXiv:1908.11103}},
  \href {https://doi.org/10.1103/PhysRevD.100.125021}
  {\path{doi:10.1103/PhysRevD.100.125021}}.

\bibitem{1911.03340}
G.~Fodor, {A review on radiation of oscillons and oscillatons}, Ph.D. thesis,
  Wigner RCP, Budapest (2019).
\newblock \href {http://arxiv.org/abs/1911.03340} {\path{arXiv:1911.03340}}.

\bibitem{2003.10899}
M.~Gleiser, M.~Krackow, {Configurational Entropic Study of the Enhanced
  Longevity in Resonant Oscillons}, Phys. Lett. B805 (2020) 135450.
\newblock \href {http://arxiv.org/abs/2003.10899} {\path{arXiv:2003.10899}},
  \href {https://doi.org/10.1016/j.physletb.2020.135450}
  {\path{doi:10.1016/j.physletb.2020.135450}}.

\bibitem{Zhang:2020bec}
H.-Y. Zhang, M.~A. Amin, E.~J. Copeland, P.~M. Saffin, K.~D. Lozanov,
  {Classical Decay Rates of Oscillons}, JCAP 2007 (2020) 055.
\newblock \href {http://arxiv.org/abs/2004.01202} {\path{arXiv:2004.01202}},
  \href {https://doi.org/10.1088/1475-7516/2020/07/055}
  {\path{doi:10.1088/1475-7516/2020/07/055}}.

\bibitem{1807.09795}
J.~Eby, K.~Mukaida, M.~Takimoto, L.~C.~R. Wijewardhana, M.~Yamada, {Classical
  nonrelativistic effective field theory and the role of gravitational
  interactions}, Phys. Rev. D99~(12) (2019) 123503.
\newblock \href {http://arxiv.org/abs/1807.09795} {\path{arXiv:1807.09795}},
  \href {https://doi.org/10.1103/PhysRevD.99.123503}
  {\path{doi:10.1103/PhysRevD.99.123503}}.

\bibitem{1609.04724}
T.~Helfer, D.~J.~E. Marsh, K.~Clough, M.~Fairbairn, E.~A. Lim, R.~Becerril,
  {Black hole formation from axion stars}, JCAP 1703 (2017) 055.
\newblock \href {http://arxiv.org/abs/1609.04724} {\path{arXiv:1609.04724}},
  \href {https://doi.org/10.1088/1475-7516/2017/03/055}
  {\path{doi:10.1088/1475-7516/2017/03/055}}.

\bibitem{1708.01344}
T.~Ikeda, C.-M. Yoo, V.~Cardoso, {Self-gravitating oscillons and new critical
  behavior}, Phys. Rev. D96~(6) (2017) 064047.
\newblock \href {http://arxiv.org/abs/1708.01344} {\path{arXiv:1708.01344}},
  \href {https://doi.org/10.1103/PhysRevD.96.064047}
  {\path{doi:10.1103/PhysRevD.96.064047}}.

\bibitem{1906.09346}
F.~Muia, M.~Cicoli, K.~Clough, F.~Pedro, F.~Quevedo, G.~P. Vacca, {The Fate of
  Dense Scalar Stars}, JCAP 1907 (2019) 044.
\newblock \href {http://arxiv.org/abs/1906.09346} {\path{arXiv:1906.09346}},
  \href {https://doi.org/10.1088/1475-7516/2019/07/044}
  {\path{doi:10.1088/1475-7516/2019/07/044}}.

\bibitem{1907.10613}
E.~Cotner, A.~Kusenko, M.~Sasaki, V.~Takhistov, {Analytic Description of
  Primordial Black Hole Formation from Scalar Field Fragmentation}, JCAP 1910
  (2019) 077.
\newblock \href {http://arxiv.org/abs/1907.10613} {\path{arXiv:1907.10613}},
  \href {https://doi.org/10.1088/1475-7516/2019/10/077}
  {\path{doi:10.1088/1475-7516/2019/10/077}}.

\bibitem{1912.09658}
X.-X. Kou, C.~Tian, S.-Y. Zhou, {Oscillon Preheating in Full General
  Relativity} (2019).
\newblock \href {http://arxiv.org/abs/1912.09658} {\path{arXiv:1912.09658}}.

\bibitem{0803.3085}
E.~Silverstein, A.~Westphal, {Monodromy in the CMB: Gravity Waves and String
  Inflation}, Phys. Rev. D78 (2008) 106003.
\newblock \href {http://arxiv.org/abs/0803.3085} {\path{arXiv:0803.3085}},
  \href {https://doi.org/10.1103/PhysRevD.78.106003}
  {\path{doi:10.1103/PhysRevD.78.106003}}.

\bibitem{McAllister:2014mpa}
L.~McAllister, E.~Silverstein, A.~Westphal, T.~Wrase, {The Powers of
  Monodromy}, JHEP 09 (2014) 123.
\newblock \href {http://arxiv.org/abs/1405.3652} {\path{arXiv:1405.3652}},
  \href {https://doi.org/10.1007/JHEP09(2014)123}
  {\path{doi:10.1007/JHEP09(2014)123}}.

\bibitem{Hu:2000ke}
W.~Hu, R.~Barkana, A.~Gruzinov, {Cold and fuzzy dark matter}, Phys. Rev. Lett.
  85 (2000) 1158--1161.
\newblock \href {http://arxiv.org/abs/astro-ph/0003365}
  {\path{arXiv:astro-ph/0003365}}, \href
  {https://doi.org/10.1103/PhysRevLett.85.1158}
  {\path{doi:10.1103/PhysRevLett.85.1158}}.

\bibitem{Marsh:2013ywa}
D.~J.~E. Marsh, J.~Silk, {A Model For Halo Formation With Axion Mixed Dark
  Matter}, Mon. Not. Roy. Astron. Soc. 437~(3) (2014) 2652--2663.
\newblock \href {http://arxiv.org/abs/1307.1705} {\path{arXiv:1307.1705}},
  \href {https://doi.org/10.1093/mnras/stt2079}
  {\path{doi:10.1093/mnras/stt2079}}.

\bibitem{Schive:2014dra}
H.-Y. Schive, T.~Chiueh, T.~Broadhurst, {Cosmic Structure as the Quantum
  Interference of a Coherent Dark Wave}, Nature Phys. 10 (2014) 496--499.
\newblock \href {http://arxiv.org/abs/1406.6586} {\path{arXiv:1406.6586}},
  \href {https://doi.org/10.1038/nphys2996} {\path{doi:10.1038/nphys2996}}.

\bibitem{Schive:2014hza}
H.-Y. Schive, M.-H. Liao, T.-P. Woo, S.-K. Wong, T.~Chiueh, T.~Broadhurst,
  W.-Y.~P. Hwang, {Understanding the Core-Halo Relation of Quantum Wave Dark
  Matter from 3D Simulations}, Phys. Rev. Lett. 113~(26) (2014) 261302.
\newblock \href {http://arxiv.org/abs/1407.7762} {\path{arXiv:1407.7762}},
  \href {https://doi.org/10.1103/PhysRevLett.113.261302}
  {\path{doi:10.1103/PhysRevLett.113.261302}}.

\bibitem{Hui:2016ltb}
L.~Hui, J.~P. Ostriker, S.~Tremaine, E.~Witten, {Ultralight scalars as
  cosmological dark matter}, Phys. Rev. D 95~(4) (2017) 043541.
\newblock \href {http://arxiv.org/abs/1610.08297} {\path{arXiv:1610.08297}},
  \href {https://doi.org/10.1103/PhysRevD.95.043541}
  {\path{doi:10.1103/PhysRevD.95.043541}}.

\bibitem{Marsh:2015wka}
D.~J.~E. Marsh, A.-R. Pop, {Axion dark matter, solitons and the
  cusp\textendash{}core problem}, Mon. Not. Roy. Astron. Soc. 451~(3) (2015)
  2479--2492.
\newblock \href {http://arxiv.org/abs/1502.03456} {\path{arXiv:1502.03456}},
  \href {https://doi.org/10.1093/mnras/stv1050}
  {\path{doi:10.1093/mnras/stv1050}}.

\bibitem{Clough:2015sqa}
K.~Clough, P.~Figueras, H.~Finkel, M.~Kunesch, E.~A. Lim, S.~Tunyasuvunakool,
  {GRChombo : Numerical Relativity with Adaptive Mesh Refinement}, Class.
  Quant. Grav. 32~(24) (2015) 245011, [Class. Quant. Grav.32,24(2015)].
\newblock \href {http://arxiv.org/abs/1503.03436} {\path{arXiv:1503.03436}},
  \href {https://doi.org/10.1088/0264-9381/32/24/245011}
  {\path{doi:10.1088/0264-9381/32/24/245011}}.

\bibitem{Michel:2018nzt}
F.~Michel, I.~G. Moss, {Relativistic collapse of axion stars}, Phys. Lett. B
  785 (2018) 9--13.
\newblock \href {http://arxiv.org/abs/1802.10085} {\path{arXiv:1802.10085}},
  \href {https://doi.org/10.1016/j.physletb.2018.07.063}
  {\path{doi:10.1016/j.physletb.2018.07.063}}.

\end{thebibliography}

\end{document}